\documentclass[a4paper,12pt]{article}
\usepackage{ifthen}
\newcommand{\remark}[1]{ }

\newcommand{\sect}[1]{Section~\ref{sec:#1}}
\newcommand{\app}[1]{Appendix~\ref{app:#1}}

\newcommand{\eqnnum}[1]{(\ref{eqn:#1})}
\newcommand{\eqn}[1]{Eq.~\eqnnum{#1}}
\newcommand{\bibnum}[1]{\cite{bib:#1}}
\newcommand{\bib}[1]{reference~\bibnum{#1}}
\newcommand{\beq}{\begin{eqnarray}}
\newcommand{\eeq}{\end{eqnarray}}
\newcommand{\beqa}{\beq \ba{rcl}}
\newcommand{\eeqa}{\ea \eeq}
\newcommand{\eqal}[3]{ \displaystyle #1 &#2& \displaystyle #3 \\ }
\newcommand{\leqal}[3]{ \displaystyle #1 &#2& \displaystyle #3 }
\newcommand{\bed}{\begin{displaymath}}
\newcommand{\eed}{\end{displaymath}}
\newcommand{\bei}{\begin{itemize}}
\newcommand{\eei}{\end{itemize}}
\newcommand{\ben}{\begin{enumerate}}
\newcommand{\een}{\end{enumerate}}

\newcommand{\bep}{\begin{picture}}
\newcommand{\eep}{\end{picture}}

\newcommand{\bef}{\begin{figure}}
\newcommand{\eef}{\end{figure}}
\newcommand{\bet}{\begin{table}}
\newcommand{\eet}{\end{table}}

\newcommand{\elab}[1]{\label{eqn:#1}}

\newcommand{\slab}[1]{\label{sec:#1}}
\newcommand{\alab}[1]{\label{app:#1}}

\remark{
\makeatletter 
         
         \@addtoreset{equation}{section}
\makeatother
}
\newcommand{\myappendix}{
\appendix
\addtocontents{toc}{\protect\newsec}}

\newcommand{\appsection}[1]{
\renewcommand{\thesection}{Appendix~\Alph{section}}
\section{#1}
\renewcommand{\thesection}{\Alph{section}}}

\newlength{\appwidth}
\makeatletter
\newcommand{\newsec}{
\renewcommand{\l@section}[2]{%
    \addpenalty{\@secpenalty}%
    \addvspace{1.0em \@plus\p@}%
    \setlength\@tempdima{1.5em}%
    \addtolength{\@tempdima}{\appwidth}%
    \begingroup
    \parindent \z@ \rightskip \@pnumwidth
    \parfillskip -\@pnumwidth
    \leavevmode \bfseries
    \advance\leftskip\@tempdima
    \hskip -\leftskip
    ##1\nobreak\hfil \nobreak\hbox to\@pnumwidth{\hss ##2}\par
   \endgroup}
}
\makeatother

\newcounter{sub}

\newcommand{\grkend}[1]{
\ifthenelse{\equal{#1}{1}}{\mu}{}
\ifthenelse{\equal{#1}{2}}{\nu}{}
\ifthenelse{\equal{#1}{3}}{\rho}{}
\ifthenelse{\equal{#1}{4}}{\sigma}{}
\ifthenelse{\equal{#1}{5}}{\epsilon}{}
\ifthenelse{\equal{#1}{6}}{\kappa}{}}
\newcommand{\grksp}[1]{
\ifthenelse{\equal{#1}{1}}{\alpha}{}
\ifthenelse{\equal{#1}{2}}{\beta}{}
\ifthenelse{\equal{#1}{3}}{\gamma}{}
\ifthenelse{\equal{#1}{4}}{\delta}{}
\ifthenelse{\equal{#1}{5}}{\epsilon}{}
\ifthenelse{\equal{#1}{6}}{\kappa}{}}

\newcommand{\scfeps}[2]{\ensuremath{\epsilon_{\grksp{#1} \grksp{#2}}     }}
\newcommand{\scfepsb}[2]{\ensuremath{\epsilon_{\dot{\grksp{#1}} \dot{\grksp{#2}} }    }}

\newcommand{\scfs}[1]{\ensuremath{S_{\grksp{#1}}     }}

\newenvironment{lbrce}{\left\{}{\right\}}
\newcommand{\lset}{ \begin{lbrce} }
\newcommand{\lsqu}{\left[}
\newcommand{\lprn}{\left(}
\newcommand{\rset}{\end{lbrce} }
\newcommand{\rsqu}{\right]}
\newcommand{\rprn}{\right)}
\newlength{\ifdota}
\newlength{\ifdotb}
\newcommand{\ifdot}[1]{
\settowidth{\ifdota}{\mbox{\ensuremath{a(#1)}}}
\ifthenelse{\lengthtest{\ifdota = \ifdotb}}{ }{ \ensuremath{\dot{#1}}}
}
\newcommand{\ifbrack}[1]{
\settowidth{\ifdota}{\mbox{\ensuremath{a(#1)}}}
\ifthenelse{\lengthtest{\ifdota = \ifdotb}}{ }{ \ensuremath{(#1)}}
}
\newcommand{\swp}[2]{\ensuremath{P_{\grksp{#1} \ifdot{\grksp{#2}}}     }}
\newcommand{\swd}{\ensuremath{D   }}
\newcommand{\swk}[3][{}]{\ensuremath{K^{\ifbrack{#1}\grksp{#2}\ifdot{\grksp{#3}}}     }}
\newcommand{\lbswk}[3][{}]{\ensuremath{K_{0}^{#1\grksp{#2}\ifdot{\grksp{#3}}}     }}
\newcommand{\mlswk}[3]{\ensuremath{K_{#3}^{\grksp{#1}\ifdot{\grksp{#2}}}     }}
\newcommand{\boldq}{\ensuremath{\mathbf{q}     }}
\newcommand{\swi}{\ensuremath{T    }}
\newcommand{\swq}[2][{}]{\ensuremath{Q_{\grksp{#2} \ifbrack{#1}}     }}
\newcommand{\swqb}[2][{}]{\ensuremath{\bar{Q}^{\ifbrack{#1}}_{\ifdot{\grksp{#2}}}     }}
\newcommand{\sws}[2][{}]{\ensuremath{S^{\grksp{#2} \ifbrack{#1}}     }}
\newcommand{\swsb}[2][{}]{\ensuremath{\bar{S}^{\ifdot{\grksp{#2}}}_{\ifbrack{#1}}     }}
\newcommand{\swr}{\ensuremath{R    }}

\newcommand{\pdv}[2]{\ensuremath{\frac{\partial #1}{\partial #2}}}
\newcommand{\sdrv}[3]{\ensuremath{\partial_{#1\grksp{#2}\ifdot{\grksp{#3}}}}}

\newcommand{\dq}[1][{}]{\ensuremath{\delta q_{#1}}}

\newcommand{\ess}[3]{\ensuremath{ s_{#1}^{ \grksp{#2} \ifdot{\grksp{#3}} }   }}
\newcommand{\kay}[2]{\ensuremath{ k_{ \grksp{#1} \ifdot{\grksp{#2}} }   }}
\newcommand{\sess}[1]{\ensuremath{s_{#1}^{ 2}   }}
\newcommand{\duess}[3]{\ensuremath{ {s_{#1\grksp{#2}}}^{  \ifdot{\grksp{#3}} }   }}
\newcommand{\udess}[3]{\ensuremath{ { {s_{#1}}^{\grksp{#2}} }_{ \ifdot{\grksp{#3}}}   }}
\newcommand{\ddess}[3]{\ensuremath{ s_{#1 \grksp{#2} \ifdot{\grksp{#3}}}   }}
\newcommand{\invess}[3]{\ensuremath{{s^{-1}_{#1}}^{ \grksp{#2} \ifdot{\grksp{#3}}}   }}
\newcommand{\duinvess}[3]{\ensuremath{ {s^{-1}_{#1\grksp{#2}}}^{  \ifdot{\grksp{#3}} }   }}

\newcommand{\ddinvess}[3]{\ensuremath{s^{-1}_{#1 \grksp{#2} \ifdot{\grksp{#3}}}   }}
\newcommand{\essdot}[2]{\ensuremath{(\ess{#1}{}{} . \ddess{#2}{}{})  }}
\newcommand{\essprd}[4]{\ensuremath{(\ess{#1}{}{} . \ddess{#2}{}{} . \ess{#3}{}{} . \ddess{#4}{}{})  }}

\newcommand{\esspair}[4]{\ensuremath{(\ess{#1}{}{} . \ddess{#2}{}{})_{\grksp{#3} \grksp{#4} }  }}
\newcommand{\sth}[2][{}]{\ensuremath{\theta_{#2}^{\ifbrack{#1}2}    }}
\newcommand{\dth}[3][{}]{\ensuremath{\theta^{\ifbrack{#1}}_{#2 \grksp{#3}}    }}
\newcommand{\uth}[3][{}]{\ensuremath{\theta^{\ifbrack{#1}\grksp{#3}}_{#2}    }}
\newcommand{\mth}[3][{}]{\ensuremath{\theta^{\ifbrack{#1}#3}_{#2}    }}
\newcommand{\Th}[1]{\ensuremath{\Theta^{\grksp{#1}}     }}
\newcommand{\Ch}[2]{\ensuremath{\bar{\chi}_{#1\ifdot{\grksp{#2}}}     }}
\newcommand{\uCh}[2]{\ensuremath{\bar{\chi}_{#1}^{\ifdot{\grksp{#2}}}     }}
\newcommand{\sgsp}[2][{}]{\ensuremath{
\bar{\Lambda}_{#2}^{\ifbrack{#1} 2}     
}}
\newcommand{\dgsp}[3][{}]{\ensuremath{{{\bar{\Lambda}_{#2\ifdot{\grksp{#3}}}}}^{\ifbrack{#1}}     }}
\newcommand{\ugsp}[3][{}]{\ensuremath{
\bar{\Lambda}_{#2}^{\ifbrack{#1}\ifdot{\grksp{#3}}}     
}}

\newcommand{\qsum}{\ensuremath{{q_{0}}}}

\newcounter{lisno}

\newcommand{\p}{\partial}
\newcommand{\ba}{\begin{array}}
\newcommand{\ea}{\end{array}}
\newcommand{\es}[1]{\ensuremath{#1}}

\newcommand{\h}{\alpha}

\newcommand{\e}{\epsilon}

\begin{document}

\title{Chiral Green's Functions in Superconformal Field Theory \vspace{-2 in}  \begin{flushright} \normalsize LTH-452 \\ KCL-MTH-99-11 \end{flushright} \vspace{2 in} }
\author{{\bf Austin Pickering\thanks{e.mail:pickring@amtp.liv.ac.uk}\ \ } \\ Dept. of Mathematical Sciences \\ University of Liverpool  \\ Liverpool \\ L69 3BX \\ \\ {\bf Peter West} \\ Dept. of Mathematics \\ King's College, London \\ WC2R 2LS}
\date{}
\maketitle

\begin{abstract}
By solving   the Ward identities in a  superconformal field theory we find the unique three-point Green's functions composed of chiral superfields for $N= 1,2,3,4$ supersymmetry. We show that the $N=1$ four-point function with $R$-charge equal to one is uniquely determined by the Ward identities up to the specification of four constants. We discuss  why chiral Green's functions above three-points, with total $R$-charge greater than $N$, are not uniquely determined.  
\end{abstract}

~
\vspace{1 in}
~

\pagebreak

\section{Introduction} 

\slab{introone}

The symmetries of  Maxwell's classical equations have played a defining role in modern physics. Their importance for relating observers which moved at constant velocity with respect to each other was realised by Lorentz, prior to the development of special relativity by Einstein in 1905. Their gauge symmetry was discovered by Weyl in the 1920's and was extended to construct the standard model in
1967, However, it is only more recently that  the true importance  of
electromagnetic duality and conformal invariance has become apparent.
In fact,  the  conformal invariance of the classical Maxwell equations was realised as
long ago as 1909 \bibnum{pcwone}. Unfortunately, quantum effects in Maxwell's
theory coupled to electrons and in all other   four-dimensional
theories which were subsequently  studied for many years, lead  to
violations of their conformal symmetries. The corresponding anomaly
is directly related to the appearance of infinities  in quantum
field theory.  Despite this,  in the 1960's and 1970's there was
 a revival of interest in four dimensional
conformal symmetry \bibnum{pcwtwo} and it was found that the two and three point
Green's functions could
be determined up to constants by conformal symmetry \bibnum{pcwthr}.
With the discovery of supersymmetry,
 examples of conformally invariant four dimensional quantum field
theories were found. The first  such  example \bibnum{pcwfou} being     the
$N=4$  Yang-Mills theory. Subsequently, it was  realised that
there were an infinite number of $N=2$ theories \bibnum{pcwfiv} and even some
$N=1$ theories \bibnum{pcwsix}.
More recently other examples of conformally invariant
supersymmetric theories have been found \bibnum{pcwsev}.

Supersymmetric theories are most naturally  formulated in terms  of
superfields, since only then is their supersymmetry manifest and,
as a result, can their quantum properties be most systematically
studied. However,  the   superfields which describe physical
quantities  are always subject to constraints.
For example, the Wess-Zumino model and the field strengths of $N=1$
and
$N=2$ Yang-Mills theory are described by chiral superfields.  In fact,
 these  constraints,
which imply that these superfields
in effect live on only a subspace of the usual Minkowski superspace,
 are directly responsible for the well known  non-renormalisation
theorems in supersymmetric theories \bibnum{pcweig}.  As a  result of the pattern
found when calculating the chiral  Green's functions in   two dimensional 
$N=2$ superconformal minimal models \bibnum{pcwnin} it
was proposed  that   the constraints on the superfields  when combined
with superconformal invariance could also lead to results  in four
dimensions which were stronger than those that were  generically found
in conformally invariant, but  non-supersymmetric theories.  The first
such result was the  realisation that the relation between the $R$-weight and the dilation weight of any chiral superfield could be used to
determine its dilation weight in a superconformal theory \bibnum{l9scfpap7}.  In
reference \bibnum{l9scfpap8} the chiral Ward identities for any $N$ were given and
it was shown  that there were no superconformal  chiral
invariants. In a subsequent  series of papers \bibnum{pcwtwe} it was also
realised  that theories that involved harmonic superfields, such as
$N=4$ Yang-Mills theory  would have very strong constraints placed on
them as a result of their superconformal invariance.

An early dicussion of three-point functions in $N=1$ superconformal theories appears in reference \bibnum{mpsetal}. In reference \bibnum{l9scfpap7}  an expression for the three-point Green's function
for
$N=1$ chiral superfields was given. Unfortunately,
this expression was not correct and
was subsequently corrected by
the authors of the present paper in the thesis of reference \bibnum{apthesis}.
Although the work in this thesis was made available to some
workers, and some of its results were reviewed in reference \bibnum{pcwfoutnb},
it is not  available to most workers in the field. A discussion of the superconformal group was given in references \bibnum{pcwfoutna}, \bibnum{conthesis} and \bibnum{pcwrig}. In this paper we
give some of the results of reference \bibnum{apthesis}
 and extend them by calculating
 the most general expression for the three-point chiral Green's function for
any $N$. The result can be succinctly summarised as
\beq
G_{3}^{(N)} &=& \lprn \frac{\sess{12}\sess{23}}{\sess{13}} \rprn^{(N-2)} \lprn \frac{\sess{12}}{\sess{13}} \rprn^{\frac{(4-N)q_3}{N} } 
\lprn \frac{\sess{23}}{\sess{13}} \rprn^{\frac{(4-N)q_1}{N}} \prod_{i=1}^{N} \sgsp[i]{123}, \elab{introgenres}
\eeq
where $\sum_i q_i = N$.

 Additionally,  we show that \bibnum{apthesis}, contrary to na\"{\i}ve expectations, this does not in general imply that the chiral Green's functions higher than three-points
are determined up to constants. In fact, as a direct consequence of  the nilpotent properties of  these Green's functions, we find that we can not uniquely determine any solution above three-points when the total $R$-charge of the Green's function, denoted by \qsum, is greater than one. However, in the particular case when $\qsum = 1$, we find that the $N=1$ four-point solution is uniquely specified up to four constants by the superconformal Ward identities and we show how to construct the appropriate solution once these constants have been specified.

\section{General Properties of Solutions}

\slab{introtwo}

As discussed in \bib{l9scfpap8}, the superconformal Ward identities for translations, dilations and  special conformal transformations acting on chiral Green's functions, $G$,   are
\beq
\swp{1}{1} \, G &=& \sum_{p=1}^{n}  \lset \partial_{\alpha \dot \alpha}  \rset  G=0, \elab{l9scwp}
\eeq
\beq
\swd \, G &=& \sum_{p=1}^{n}    \lset s ^{\alpha \dot \alpha } \partial_{\alpha \dot \alpha} + {1\over 2} \theta ^{\alpha j} \partial_{\alpha j } + q{(4-N)\over N}  \rset  G=0, \elab{l9scwd}
\eeq
\beq
\swk{2}{2} \, G &=& \sum_{p=1}^{n}    \lset s ^{\alpha \dot \beta } s ^{\beta \dot \alpha } \partial_{\alpha \dot \alpha} +s ^{\alpha \dot \beta } \theta ^{\beta j} \partial_{\alpha j } + q{(4-N)\over N}s^{\beta \dot \beta}   \rset G=0, \elab{l9scwk}
\eeq
For supersymmetry, they are
\beq 
\swq[i]{1} \, G &=& \sum_{p=1}^{n}   \lset {\partial_{\alpha i} }\rset  G=0, \elab{l9scwq}
\eeq
\beq 
\swqb[i]{1} \, G &=& \sum_{p=1}^{n}   \lset \theta^{\alpha i} {\partial_{\alpha \dot \alpha} } \rset G =0. \elab{l9scwqb}
\eeq
For the internal symmetry, with traceless parameter $ E_j^{\ i}$, we
have the corresponding Ward identity
\beq
\swi \, G &=&  \sum_{p=1}^{n}  \lset \theta ^{\alpha j} E_j^{\ i} \partial _{\alpha i} \rset G=0,  \elab{l9scwi}
\eeq
and finally those for $S$-supersymmetry are given by
\beq 
\swsb[i]{1} \, G &=& \sum_{p=1}^{n}     \lset s ^{\beta \dot \alpha  }\partial _{ \beta i}  \rset G=0,  \elab{l9scwsb}
\eeq
\beq
\sws[i]{2} \, G &=& \sum_{p=1}^{n}  \lset s ^{\beta \dot\alpha } \theta ^{\alpha i}\partial _{\alpha \dot \alpha} - \theta ^{\beta i} \theta ^{\alpha j}\partial_{\alpha j} + q{(4-N)\over N}\theta ^{\beta i}  \rset G=0.  \elab{l9scws}
\eeq
In the above equations the sum is over $p$, however, this index is
suppressed on the coordinates and on $q$. We have used the shorthand notation
$\partial_{\alpha \dot \alpha}={\partial \over \partial s^{\alpha
\dot \alpha}}$ and
$\partial _{ \alpha i}= {\partial \over \partial \theta
^{\alpha  i}}$.
For $N\neq 4$ we also have $R$ symmetry, with the corresponding
Ward identity
\beq 
\swr \, G &=& \sum_{p=1}^{n}   \lset\theta ^{\alpha j} {\partial_{\alpha j}} -2q \rset G=0.  \elab{l9scwr}
\eeq

The operators \[ \lset \swp{1}{1}, \swk{1}{1}, \swd, \swq{1}, \swqb{1}, \sws{1}, \swsb{1}, \swr \rset , \] in the above, obey the superconformal algebra. 
The variable \ess{p}{1}{1} is a chiral variable and takes the form 
\beq
\ess{}{1}{1} &=& x^{\alpha \dot\alpha} - \frac{i}{2} {\uth{}{1}}^{j} \bar{\theta}^{\dot\alpha}_{j}.
\eeq

To begin with, we consider the case $N=1$, where \eqn{l9scwi} is trivially realised since $E$  is  zero. 
It is well known that the solution of the constraint in \eqn{l9scwp} implies that the Green's functions are functions of the differences \ess{pq}{1}{2} only,  where $q=p+1$. This is easy to see if we consider our independent variables to be the differences \ess{pq}{1}{2}, where $q=p+1$, along with the sum  
\beq
\ess{0}{1}{2} &\equiv& \sum_{p}^{n}   \ess{p}{1}{2} .
\eeq
Clearly, any function of \ess{p}{1}{1} can be written in terms of these variables instead.
It follows from the chain rule that, for any Green's function $G$ obeying \eqn{l9scwp}, we may write 
\beq
\pdv{G}{\ess{0}{1}{1}} 
&=& \sum_{p}^{n}   \pdv{G}{\ess{p}{1}{1}} 
\, \, \, \, \, = \, \, \, \, \,  0,
\eeq
and the result follows.

A similar argument shows that the same is true for the $\uth{p}{1}$ variables. Defining 
\beq
\Th{1} &=& \sum_{p}^{n}   \uth{p}{1} ,
\eeq
 we see that 
\beq
\pdv{G}{\Th{1}} 
&=& \sum_{p}^{n}   \pdv{G}{\uth{p}{1}}  
\, \, \, \, \, = \, \, \, \, \,  0,
\eeq
and thus $G$ is independent of $\Th{1}$ from  \eqn{l9scwq}. 

Given these simplifications, one might wonder whether any of the other operators in the algebra may be expressed as the derivative of a single variable by a suitable choice of independent variables. In particular, we consider \eqn{l9scwsb}, as it has a very simple form. We use the fact that $G$ is a function of \ess{p,p+1}{1}{1} and \uth{p,p+1}{1} only, and as a result,  we may use the chain rule to write \swsb{1} as 
\beq
\swsb{1}  &=& 
\remark{\sum_{p=1}^{n-1}   \ess{p}{2}{1} \pdv{\uth{p,p+1}{3}}{\uth{p}{2}} \pdv{G}{\uth{p,p+1}{3}} + \sum_{p=2}^{n } \ess{p}{2}{1} \pdv{\uth{p-1,p}{3}}{\uth{p}{2}} \pdv{G}{\uth{p-1,p}{3}} 
\nonumber \\ &=& \sum_{p=1}^{n-1}   \ess{p}{2}{1} \pdv{G}{\uth{p,p+1}{2}} - \sum_{p=2}^{n } \ess{p}{2}{1} \pdv{G}{\uth{p-1,p}{2}} 
\nonumber \\ &=& }
\sum_{p=q-1}^{n-1} \ess{pq}{2}{1} \pdv{G}{\uth{pq}{2}} . \elab{l9swsb2}
\eeq
However, we wish to go further and write it as 
\beq
\swsb{1} \, = \, \pdv{G}{\Ch{0}{1}} &=& \sum_{p=q-1}^{n-1} \pdv{\uth{pq}{2}}{\Ch{pq}{1}} \pdv{G}{\uth{pq}{2}} ,
\eeq
for some variable \Ch{0}{1}, which must be a function of \ess{p,p+1}{1}{1} and \uth{p,p+1}{1}, where we also have 
\beq
\Ch{0}{1} &=& \sum_{p=1}^{n-1} \Ch{p, p+1}{1} \elab{l9chidef}
\eeq
for some variables \Ch{pq}{1}.
Comparing this with \eqn{l9swsb2} we find  
\beq
\pdv{\uth{pq}{2}}{\Ch{pq}{1}} &=& \ess{pq}{2}{1},  \, \, \,\, \, \,\, \, \, p = q-1,
\eeq
which implies 
\beq
\Ch{pq}{1} &=&  \uth{pq}{2} \ddinvess{pq}{2}{1},  \elab{l9chipqdef}
\eeq
and therefore 
\beq
\Ch{0}{1} &=& \sum_{p=q-1}^{n-1} \uth{pq}{2} \ddinvess{pq}{2}{1}.  \elab{l9spindep}
\eeq
Clearly, one can write any function of \ess{p,p+1}{1}{1} and \uth{p,p+1}{1} in terms \ess{p,p+1}{1}{1}, \Ch{0}{1} and \dgsp{p, p+1, p+2}{1}, defined by 
\beq
\ugsp{pqr}{1} &=& \uCh{pq}{1} - \uCh{qr}{1}.
\eeq  
From \eqn{l9spindep}, $G$ is independent of \Ch{0}{1} and must therefore be a function of the remaining independent variables, namely the \dgsp{pqr}{1} and the  \ess{p,p+1}{1}{1}. 
In summary, any arbitrary function, $G( \ugsp{pqr}{1}, \ess{pq}{1}{1} )$, obeys Equations \eqnnum{l9scwp}, \eqnnum{l9scwq} and \eqnnum{l9scwsb}, which leaves the Ward identities \eqnnum{l9scwd}, \eqnnum{l9scwk}, \eqnnum{l9scwqb}, \eqnnum{l9scws} and \eqnnum{l9scwr} to be solved.

\section{A Particular Three-Point Solution}

\slab{partic}

In the case of the three-point function, we see immediately from the above that there is only one independent spinor, namely \ugsp{123}{1}. We can see that a solution proportional to \ugsp{123}{1} alone is not possible even if we consider non-scalar solutions. The general form of such a solution would have to be 
\beq
G_{3}' &=& \ugsp{123}{1} \, h_{\ifdot{\grksp{1}} \grksp{2}}(\ess{pq}{}{}).
\eeq
The action of \swqb{3} on this function yields  two linearly independent terms, which can each be set to zero using the Ward identity in \eqnnum{l9scwqb}. These are : 
\beq
\sth{12} \, \, \Rightarrow \,\,\,  \sdrv{12}{3}{3} \lprn \invess{12}{3}{1} h_{\ifdot{\grksp{1}} \grksp{2}} \rprn &=& 0, \\ 
\sth{23} \, \, \Rightarrow \,\,\,  \sdrv{23}{3}{3} \lprn \invess{23}{3}{1} h_{\ifdot{\grksp{1}} \grksp{2}} \rprn &=& 0,
\eeq
from which it is clear that $h_{\ifdot{\grksp{1}} \grksp{2}} = 0$, and thus there is no solution of this form.

For now, we restrict our attention to scalar solutions and deduce that the scalar three-point Green's function must be of the form 
\beq
G_{3} &=& \frac{1}{2} f(\ess{12}{1}{1}, \ess{23}{1}{1}) \, \sgsp{123} \elab{l93ptgf},
\eeq
and  the dependence on \uth{p}{1} is completely fixed. We must now determine the scalar function $f(\ess{12}{1}{1}, \ess{23}{1}{1})$ such that the remaining Ward identities are obeyed.  

First note that the Ward identity of \eqn{l9scwd} involves the operator $\ess{}{1}{1} \partial_{\alpha \dot\alpha}$ which merely counts the overall power of \ess{pq}{1}{1} in $G$. The Ward identity of \eqn{l9scwr} involves the operator  $\uth{j}{1} \partial_{\alpha j}$ which does the same for $\uth{j}{1}$ and thus \swr-symmetry fixes the value of 
\beq
\qsum &\equiv& \sum_{p=1}^{n} q_{p}.  \elab{l9qsum}
\eeq 
In this case,  \eqn{l93ptgf} shows that  
$\qsum = 1$ is the only possibility. 
One can then see by inspection that the dilation operator $\swd$ constrains $f$  to be of degree $-2$ in \ess{pq}{1}{1}.

Consider next the Ward identity of \eqn{l9scwqb}. This includes the 
action of the operator  \swqb{1} on $G_{3}$, which gives 
\beq
\swqb{1} \left[ \frac{1}{2} f \sgsp{123} \right] &=& -  f \dgsp{123}{1} \left[ \frac{\sth{12}}{\sess{12}} - \frac{\sth{23}}{\sess{23}} \right] +   \frac{1}{2} \sgsp{123} \, \swqb{1} f   \nonumber \\ 
&=& \frac{2 f}{\sess{12}\sess{23}} \left[ \sth{12} \uth{23}{1} \ddess{23}{1}{1} - \sth{23} \uth{12}{1} \ddess{12}{1}{1} \right] +   \frac{1}{2} \sgsp{123} \, \swqb{1} f. \elab{l9qbong3}
\eeq
We note that any scalar function of \ess{12}{1}{1} and \ess{23}{1}{1} can be written in terms of the three independent variables 
\beq
a \, = \,  \sess{12}, \, \, 
& b \, = \,  \essdot{12}{23},& \, \, 
c \, = \,  \sess{23},
\eeq
where we have used the shorthand notation 
\beq
\essdot{12}{23} &=& \ess{12}{1}{1}  \ddess{23}{1}{1}.
\eeq
This follows from the equations \eqnnum{l9spsqu} and \eqnnum{l9spswap} given in \app{notapp}, which can be used to reduce any scalar expression at 3 points to a function of $a$, $b$ and $c$.
Using this we can rewrite \eqn{l9qbong3} as 
\beq
\! \! \! \! \! \! \! \! \! \! \! \! \swqb{1} \left[ \frac{1}{2} f \sgsp{123} \right] &=& \frac{(\sth{12} \uth{23}{1}  \ddess{12}{1}{1} + \sth{23} \uth{12}{1}  \ddess{23}{1}{1}) }{\sess{12}\sess{23}} \left[  1 + a \pdv{}{a} + b \pdv{}{b} + c \pdv{}{c} \right] f
\eeq
This implies that $f$ is a function of degree $-1$ in $a$, $b$, $c$ (i.e. degree $-2$ in \ess{12}{}{} and \ess{23}{}{}), which we know already from the dilation operator \swd. So, in this case \swqb{1} does not give any new constraints on $f$. 

Without loss of generality, the function $f(a,b,c)$ may be considered as an arbitrary function of degree $0$ multiplied by any particular function of degree $-1$ in its arguments, with numerator $1$. Let us choose this function to be $1/\sess{13}$. Any degree zero function is in general a function of two independent ratios of $a$, $b$ and $c$, such as $(a+ 2 b + c)/a$ and $c/a$. Thus, we may write $G_{3}$ as
\beq
G_{3} &=& \frac{1}{2} f' \left(\sess{13}/\sess{23} , \sess{12}/\sess{23}\right) \,  \frac{\sgsp{123}}{{\ess{13}{}{}}^{2}} \elab{l93ptgf2},
\eeq
which is the most general solution to Equations \eqnnum{l9scwp}, \eqnnum{l9scwd}, 
\eqnnum{l9scwq}, 
\eqnnum{l9scwqb}, \eqnnum{l9scwsb} and 
\eqnnum{l9scwr}, for a scalar three point function.

We will find the most general chiral Green's 
function in \sect{uniq}, however, as a step in this direction 
we will  now show that we can choose  $f'=1$, and  show by explicit calculation that $G_{3}$ is also a solution to \eqn{l9scws}, and so all the superconformal Ward identities. The full calculation is too long to reproduce here, however, we note that whilst $G$ has to be a function of the differences in the coordinates, i.e. \ess{p,p+1}{1}{1} and \uth{p,p+1}{1}, 
the Ward identity of \eqn{l9scws} involves $\sws{1} \, G$, which is not. 

$\sws{1} \, G_{3}$ can therefore be expressed in terms of coefficients of \uth{p}{1} which are either of the form $\sth{p}\uth{q}{1}$ or $\uth{1}{1} \uth{2}{2} \uth{3}{3}$. It is the latter case which gives most difficulty, so we shall only discuss the coefficient of $\sth{1}\uth{3}{1}$ here. 
Explicit calculation reveals that 
\beq
\scfs{2} \lsqu \frac{\sgsp{123}}{2 \sess{13}} \rsqu &=& \frac{2\sth{1}\uth{3}{1} \ess{3}{2}{1} \ddess{13}{1}{1}}{\sess{12}{\ess{13}{}{}}^{4}} 
- \frac{4\sth{1}\uth{3}{4} \ess{1}{2}{1} \ddess{23}{1}{1}}{\sess{12}\sess{23}\sess{13}} \nonumber \\ & & 
+ \frac{4\sth{1}\uth{3}{4} \ess{1}{2}{1} \ddess{12}{1}{1} \ess{12}{1}{3} \ddess{23}{3}{4}}{{\ess{12}{}{}}^{4}\sess{23}\sess{13}} 
+ \frac{4\sth{1}\uth{3}{4} \ess{1}{2}{1} \ddess{13}{1}{1} \ess{12}{1}{3} \ddess{13}{3}{4}}{\sess{12}\sess{23}{\ess{13}{}{}}^{4}} 
\nonumber \\ & & 
+ (1 - 3q_{1})\frac{2\sth{1}\uth{3}{3}  \ess{12}{2}{3} \ddess{23}{3}{3}}{\sess{12}\sess{23}\sess{13}} 
+ \frac{3q_{3}\sth{1}\uth{3}{2}  }{\sess{12}\sess{13}} 
+ \ldots, \elab{l9song3}
\eeq
where the dots denote other linearly independent terms. After some rearrangement, using in particular Equations \eqnnum{l9spsqu} and \eqnnum{l9spswap}, we obtain   
\beq
\scfs{2} \lsqu \frac{\sgsp{123}}{2 \sess{13}} \rsqu &=& (3q_{3} -1) \frac{\sth{1}\uth{3}{2}  }{\sess{12}\sess{13}} 
-  (3q_{1} -1) \frac{2\sth{1}\uth{3}{3}  \ess{12}{2}{3} \ddess{23}{3}{3}}{\sess{12}\sess{23}\sess{13}} + \ldots.
\eeq
We know that $\qsum = 1$, so it follows that  \eqn{l9scws} is only valid when 
\beq
q_{1} \,\, = \,\, q_{2} &=& q_{3} \,\, = \,\, \frac{1}{3}. \elab{l9qvals}
\eeq

Since one can show that \[ \frac{\sgsp{123}}{2 \sess{13}}, \] is invariant under cyclic permutation of 1, 2, and 3, this result can be extended to all coefficients of the form $\sth{p}\uth{q}{1}$. We note that the operators in the algebra are trivially cyclic invariants, given \eqn{l9qvals}. Thus, the coefficient of $\sth{1}\uth{3}{1}$ in $\sws{1} G_{3}$ is the same as the coefficient of $\sth{2}\uth{1}{1}$ and $\sth{3}\uth{2}{1}$, and the result follows. It remains to prove the corresponding result for the coefficient of $\uth{1}{1} \uth{2}{2} \uth{3}{3}$, which has been done, but is very laborious and yields nothing new. Once again,  \eqn{l9qvals} must apply for the result to vanish.

From the superconformal algebra, we know that \swk{1}{1} is related to the anticommutator of the special supersymmetry generators, \sws{1} and \swsb{1}, so that Equations \eqnnum{l9scwsb} and \eqnnum{l9scws} together imply \eqn{l9scwk}. Thus,
\beq
G_3^{0} &\equiv& \frac{\sgsp{123}}{2 \sess{13} } \elab{l9partsol}
\eeq
 is a solution to all of the $N=1$ Ward identities, given \eqn{l9qvals}. 
Recall that we arbitrarily chose the function $f'$ to be $1$  when proving 
that the superconformal Ward identities are satisfied.  
Of course for a specific set of  $R$ charges, and so dilation weights, the three-point function  must be unique as a consequence of 
 the standard results of ordinary conformal field theory. As such, for the charges of \eqn{l9qvals} this result is the unique result. In \sect{uniq}, and 
using the results of \sect{partic},  we will 
give a superspace proof of this fact and then find the unique three-point chiral Green's function for all possible $R$ charges.

\section{Conditions for Uniqueness of Green's Functions}

\slab{uniq}

 A standard   argument to establish the uniqueness of Greens functions  goes as follows. Consider two Green's functions, $g_{1}$ and $g_{2}$. Their ratio, $r$, will satisfy all the Ward identities with no isotropy transformations and 
so will be an invariant. If however, one can prove that there are no invariants then one can establish the uniqueness. If we were to apply this argument for 
the case of chiral Green's functions considered in this paper, then we would find that \eqn{l9scwr} implies $r$ is independent of \uth{pq}{1}. From  this, it follows that  \eqn{l9scwqb} implies $r$ must be independent of \ess{pq}{1}{1} as well, and hence simply a constant. This might, at first sight, appear to suggest that all the chiral Green's functions were unique. Such an argument was suggested in the first version of \bib{l9scfpap8}.

 In fact, \eqn{l9scwr} implies that $r$ is of degree zero in \uth{pq}{1},  but the Green's functions are proportional to some power of \uth{pq}{1} 
and so are nilpotent. It is of course not correct to divide by nilpotent 
quantities. Nonetheless, one might hope 
 that one could in effect still arrive at the correct result. However, we 
 note  that the above argument does not require $S$-supersymmetry and  we have already shown by explicit construction, that the solution to the Ward identities in the absence of \eqnnum{l9scws} is not unique ( even at three points). Therefore the above uniqueness argument  must be incorrect. 

To see why, we should ask the related question: ``If $g_{1}$ is a Green's function, does there exist another of the form $g_{2} = r(\ess{pq}{1}{1}) g_{1}$ ?'' This is equivalent to studying the ratio $r$, except that no division has occurred. $r$ has to be independent of \uth{pq}{1} or otherwise $g_2$ will not satisfy Ward Identities with the same value $q_0$ as $g_1$, which it must do by hypothesis. 

As usual, we deduce from \eqn{l9scwp} that $r$ is a function of the differences, \ess{p,p+1}{1}{1}, and from \eqn{l9scwd}, that $r$ must be of degree zero.  Equations \eqnnum{l9scwq}, \eqnnum{l9scwsb} and \eqnnum{l9scwr} are trivially realised on $g_{2}$. From \eqn{l9scwqb}, we deduce 
\beq
\swqb{1} g_{2} &=& g_{1} \, \swqb{1} r  \,\,\,\,\,\, = \,\,\,\,\,\, 0.
\eeq
This does not imply that $\swqb{1} r = 0$, because $g_{1}$ is a function of Grassmann odd variables. Considering our explicit form of $G_{3}$, we see that it is the square of the spinor \ugsp{123}{1}. If $\swqb{1} r$ was proportional to \ugsp{123}{1} then $g_{2}$ could obey \eqn{l9scwqb}.
However, $\swqb{1} r$ would be non-zero and thus $r$ would be dependent on \ess{pq}{1}{1}, violating the uniqueness argument. 

By definition, we  take  $r$ to be independent of $\theta$ and to be a scalar, 
and the dilation Ward identity implies that it is of  of degree $0$ in \ess{p, p+1}{}{}. We now  show that, when acted on by \swqb{1}, any scalar function of degree zero in \ess{pq}{}{} becomes a function of \dgsp{pqr}{1}. 

To see this, note that any scalar function of \ess{p,p+1}{1}{1} can be written in terms of \esspair{p,p+1}{q,q+1}{1}{2}. This is because a scalar is a trace of a product  of \ess{p, p+1}{1}{1} and, in order to take the trace of such a term, one must have an even number of \ess{p, p+1}{1}{1}. For example, one can write 
\beqa
\eqal{
\sess{pq}  }{=}{ \epsilon^{\grksp{2}\grksp{1}} \esspair{pq}{pq}{1}{2}}
\eqal{
\essdot{pq}{rs}  }{=}{ \epsilon^{\grksp{2}\grksp{1}} \esspair{pq}{rs}{1}{2} }
\leqal{
(s_{p}.s_{q}.s_{r}.s_{t}) }{=}{ \epsilon^{\grksp{2}\grksp{1}} \epsilon^{\grksp{4}\grksp{3}} \esspair{p}{q}{1}{3}\esspair{r}{t}{4}{2}
}
\eeqa
Hence every scalar can be written as the trace of a product of terms of the form \esspair{p,p+1}{q,q+1}{1}{2}. 
Furthermore, a scalar expression of degree 0 can always be written in terms of expressions of the form :
\[ \frac{\esspair{p,p+1}{q,q+1}{1}{2}}{\sess{p,p+1}}. \]
Therefore, all we need to do is establish the required result for all expressions of the form 
\[ \frac{\esspair{p,p+1}{q,q+1}{1}{2}}{\sess{p,p+1}} \]
and it automatically follows for all scalars by using Leibniz rule for first order linear differential operators. 

Direct calculation shows that  
\beq
\swqb{1}  \lsqu \frac{\esspair{pq}{rs}{1}{2}}{\sess{pq}} \rsqu &=& -  \lprn \ugsp{pqr}{2} + \ugsp{qrs}{2} \rprn \lprn 
\frac{\ddess{pq}{1}{1} \ddess{rs}{2}{2}}{\sess{pq}} \rprn,
\eeq
and thus any function of degree zero is a function of terms of the form \dgsp{pqr}{1} when acted on by \swqb{1}. Using the relations given in \app{notapp} we can see that \dgsp{pqr}{1} can always be written in terms of the basis set of functions \dgsp{a, (a+1), (a+2)}{1}. This means that $\swqb{1} k$ can in principle be non-zero and the solution to the Ward identities, excluding \eqn{l9scws} and \eqn{l9scwk}, is not necessarily unique. In particular, at three points $\swqb{1} r$ is simply proportional to \dgsp{123}{1}, and therefore $G_3 \, \swqb{1} r$ vanishes for any $r$ of degree zero, not just constant values.

In order to pursue the question of uniqueness, we must therefore consider the effect of either \swk{1}{1} or \sws{1} on a given Green's function, $g_{1}$. These operators are not independent, as seen from the algebra, and thus we need to consider only one of them. We choose \swk{1}{1} for reasons which will become clear below. 
\swk{1}{1} is the sum of a  part, \lbswk{1}{1}, which is first order in differential operators and so obeys the Leibniz rule and a multiplicative operator, \mlswk{1}{1}{q}, which contains the isotropy group action. These two parts are given by 
\beq
\lbswk{2}{2} G &=& \sum_{p}^{n}    \lset s ^{\alpha \dot \beta } s ^{\beta \dot \alpha } \partial_{\alpha \dot \alpha} +s ^{\alpha \dot \beta } \theta ^{\beta j} \partial_{\alpha j } \rset G \\ \mlswk{2}{2}{q} G &=& \sum_{p}^{n}   \lset q {(4-N)\over N}s^{\beta \dot \beta}   \rset G
\eeq
Recall that the $p$ index is suppressed on the coordinates and $q$ in the above, and they should properly be written $q_{p}$, etc. We define 
\beq
\remark{\qsum &=& \sum_{p=1}^{n} q_{p} \\}
\boldq &=& (q_{1}, q_{2}, q_{3}, \ldots, q_{n})
\eeq
and once again we consider 
\beq
g_{2} &=& r(\ess{pq}{1}{1}) \, g_{1},
\eeq
 where $r$ is degree $0$ in \ess{pq}{1}{1}, as explained above. It follows that 
\beq
\left . \swk{1}{1} g_{2} \right|_{\boldq = \boldq_2} &=& r \, \lbswk{1}{1} g_{1} + g_{1} \, \lbswk{1}{1} r + \mlswk{1}{1}{\boldq_{2}} (r g_{1}),
\eeq
for some $\boldq_{2}$, which is by definition the $R$ weight of the Green's function $g_2$. 
Defining $\boldq_1$ to be the $R$ weight of the Green's function $g_1$, we have  the equation 
\beq
\swk{1}{1} g_{1} &=&  \lbswk{1}{1} g_{1} + \mlswk{1}{1}{\boldq_{1}}  g_{1} \nonumber \\ &=& 0,
\eeq
and thus, writing 
\beq 
\boldq_{2} &=& \boldq_{1} + \delta\boldq
\eeq
we deduce that, for $g_{2}$ to be a Green's function, we must have 
\beq
\swk{1}{1} g_{2} &=&  g_{1} \lprn \lbswk{1}{1} r + \mlswk{1}{1}{\delta\boldq} r  \rprn \nonumber \\ &=& 0. \elab{l9g2isgrn}
\eeq

We note that unlike $\swqb{1} r$, $\swk{1}{1} r$ is independent of \uth{pq}{1} 
since  $r$ is independent of \uth{pq}{1} and so the $\theta $ dependent terms in $K^{\alpha \dot \alpha}$ do not act. It is for this reason that 
 we chose to investigate the action of \swk{1}{1} in contrast to \sws{1}. As a result, we may demand 
\beq
 \lbswk{1}{1} r + \mlswk{1}{1}{\delta\boldq} r &=& 0,
\eeq
from \eqn{l9g2isgrn}, irrespective of the nilpotence of $g_1$.
If $\delta\boldq = 0$, this is equivalent to demanding that $r$ be an ordinary  conformal invariant in \ess{pq}{1}{1}. If $\delta\boldq \neq 0$, then strictly, the two Green's functions are solutions to {\em different} Ward identities, since $\boldq_1$ and $\boldq_2$ are distinct. 
Considering the action of \lbswk{1}{1} on \sess{pq}, we find 
\beq
\lbswk{1}{1} \lprn \sess{pq} \rprn &=& \sess{pq} \lprn \ess{p}{}{} + \ess{q}{}{} \rprn^{\grksp{1}\ifdot{\grksp{1}}}, \elab{l9cononssqu}
\eeq
which is of the form of \mlswk{1}{1}{\boldq} (\sess{pq}), for a
 choice 
of \boldq\ whose values are associated with the legs $p$ and $q$. In contrast, when acting on a sum of such terms, we find that 
\beq
\lbswk{1}{1}  \lprn \sess{pq} + \sess{rs} \rprn  &=& \sess{pq} \lprn \ess{p}{}{} + \ess{q}{}{} \rprn^{\grksp{1}\ifdot{\grksp{1}}} + \sess{rs} \lprn \ess{r}{}{} + \ess{s}{}{} \rprn^{\grksp{1}\ifdot{\grksp{1}}},
\eeq
which can never be generated by  \mlswk{1}{1}{\boldq} acting on a scalar function. The same is true of any function involving the sum of two or more distinct \sess{pq}. In particular, 
\beq
\essdot{pq}{rt} &=& \sess{pt} + \sess{qr} - \sess{pr} - \sess{qt} \elab{essdotexp}
\eeq
implies that 
\beq
\swk{1}{1} \essdot{pq}{rt} &\neq& 0, \elab{l9kpair}
\eeq
from which it follows that (see \app{notapp} for notation)
\beq
\swk{1}{1} \essprd{1}{2}{3}{4} &\neq& 0
\eeq
since otherwise, choosing $\ess{3}{}{} = \ess{4}{}{}$, would give a contradiction with \eqn{l9kpair}. This accounts for all scalar expressions and we must therefore construct the  function, $r$, from products of \sess{pq}, and ensure that it is of degree zero. For such  functions, $\lbswk{1}{1} r$ is of the form $\mlswk{1}{1}{\delta \boldq} r$, and thus we may find a non-vanishing $r$ such that 
\beq 
\swk{1}{1} r &=& \lbswk{1}{1} r + \mlswk{1}{1}{\delta\boldq} r \nonumber \\ &=& 0 ,
\eeq
for a suitable choice of $\delta\boldq$. 

If we restrict our attention to the explicitly known three-point solution, given in Equations \eqnnum{l9partsol} and \eqnnum{l9qvals}, then we see that since no three-point purely conformal invariant exists, 
it must be unique once the $R$ charges $\boldq$ of the chiral Greens function are specified. However, there are an infinite number of solutions which have distinct $\boldq$. We can generate a new Green's function from an existing one simply by multiplying by a degree zero function of \sess{pq}, where there is no restriction on $p$ or $q$. At three-points, the new Green's function is a solution to a 
 set of Ward Identities with the appropriate $R$ charges. Thus, uniqueness survives at the three-point level, but only by virtue of the standard uniqueness of any conformal three-point function. It is is not due to the chirality of the Green's function. 

In contrast, one can generate a four-point solution by multiplying together two three-point solutions, with different \ugsp{pqr}{1}. We have 
\beq
G_{4} &=& r(\sess{pq}) \, \frac{\sgsp{123}}{2\sess{13}} \frac{\sgsp{234}}{\sess{24}}   ,
\eeq
for some choice of $\boldq$. This follows from the three point results for the Leibniz parts of the differential operators, and the remaining terms coming from the isotropy group action can be made to vanish by choosing \boldq\ suitably.  In this case,  $\qsum = 2$. For the particular case of $r=1$, we deduce from our knowledge of the three-point solution that 
\beq
2 q_{1} \, = \, q_{2} &=& q_{3} \, = \, 2 q_{4} \, = \, \frac{2}{3}. \elab{l9qvals4}
\eeq
However, as seen from the above discussion, uniqueness depends on the existence of an ordinary  conformal  $n$-point function, i.e. one which obeys $\lbswk{1}{1} r  = 0$. It is well known from ordinary conformal theory that such invariants exist, e.g. the independent cross ratios 
\beq
u \,\, \, \, = \,\,\,\, \frac{\sess{12}\sess{34}}{\sess{13}\sess{24}}, \,\,\,\,\,\,\,\,\, v \,\, \, \, = \,\,\,\, \frac{\sess{23}\sess{14}}{\sess{13}\sess{24}} \elab{uvdefs}.
\eeq
The fact that these are conformal invariants is easily seen from \eqn{l9cononssqu}. 

If $r$ is a function of $u, v$, then  \eqn{l9g2isgrn} is valid when $\delta\boldq = 0$. Thus, for four-points and above, there exist distinct Green's functions, $g_{1}$ and $g_{2}$, which are solutions to precisely the same Ward identities, and thus they are not unique. Given \eqn{l9qvals4}, we may write the corresponding four point solution as 
\beq
G_{4} &=& r(u,v) \, \frac{\sgsp{123}}{2\sess{13}} \frac{\sgsp{234}}{\sess{24}}   ,
\eeq
where $r(u,v)$ is completely arbitrary. 

This result can be traced directly to the nilpotence of 
$\ugsp{123}{1}$ and $\ugsp{234}{1}$. In particular, when either of these spinors is raised to the third power they vanish. Thus, one might attempt to find unique solutions, for a given \qsum,  by restricting the value of \qsum\ to $1$, and thus ruling out the possibility of this effect. 
Whilst such solutions may well exist, uniqueness is still not guaranteed.  In the case where $\qsum=1$, the solution may be of the form $\bar{\Pi}^{2}$, where $\bar{\Pi}_{\ifdot{\grksp{1}}}$ is some linear combination of the \dgsp{pqr}{1}. If $\bar{\Pi}_{\ifdot{\grksp{1}}}$ were found to be equal to $\swqb{1} h$, where $h$ was some function of the cross ratios, then we could generate new Green's functions from existing ones by multiplying them by arbitrary functions of $h$. 

We shall explore this idea later on in this paper. For the moment, we return to the three point function to find its explicit form for any given \boldq.

\section{The Full $N=1$ Three-Point Solution}

\slab{full}

We know already that \eqn{l9partsol} defines a solution to the Ward Identities in the special case where : 
\beq
\boldq &=& \boldq_1 \,\,\,\,\, = \,\,\,\,\, \lprn \frac{1}{3}, \frac{1}{3}, \frac{1}{3} \rprn
\eeq
We also know from our discussion above that we can generate a new solution to {\em different} Ward Identities, which have $\boldq = \boldq_1 + \delta\boldq$, by multiplying $G_{3}^{0}$ by any degree zero function of \ess{pq}{}{} which obeys 
\beq
\left . \swk{1}{1} r \, \right|_{\boldq = \delta\boldq}&=& 0.
\eeq
We have also seen that the only solutions to such an equation must be in the form of a sum of products of \sess{pq}, so that in general 
\beq
r &=& \sum_{a,b} c_{ab} \lprn \sess{12}\rprn^a \lprn \sess{23}\rprn^b \lprn \sess{13} \rprn^{(-a-b)}, 
\eeq
for some constants $c_{ab}$.
Acting on this with \swk{1}{1} and setting the coefficients of the linearly independent terms to zero, we see that for each term individually, we obtain 
\beq
-a + 3 \dq[3] &=& 0,  \nonumber \\
a + b + 3 \dq[2] &=& 0, \\
-b + 3 \dq[1] &=& 0. \nonumber 
\eeq
We can solve this for $a$ and $b$, given $\delta\boldq$, and thus we obtain the general form of the scalar three-point function, as
\beq
G_{3} &=&  \lprn \frac{\sess{12}}{\sess{13}} \rprn^{3q_3} 
\lprn \frac{\sess{23}}{\sess{13}} \rprn^{3q_1} 
\lprn \frac{\sess{13}}{\sess{12}\sess{23}} \rprn 
\sgsp{123} \elab{l9fullsol}
\eeq
up to a multiplicative constant. From the above arguments concerning the allowed form of $r$, it also follows that the only way to generalise this to include non-scalar three-point functions is to multiply $G_{3}$ by a constant tensor, such as $\scfeps{1}{2}$, otherwise \swk{1}{1} will not vanish.  

\section{The Complete Three-Point Solution in Extended Supersymmetry}

\slab{exten}

\subsection{$N=2$}

We can generalise the discussion of three-point solutions to situations where $N>1$. For example, when $N=2$ we have twice as many spinors, i.e. \dgsp[1]{123}{} and \dgsp[2]{123}{}. 

One might suppose that we could construct a new type of solution with $\qsum = 1$, but such a solution would have to take the form 
\beq
\left . G_{3}^{(2)} \right|_{\qsum = 1} &=&  \ugsp[1]{123}{1} f_{\ifdot{\grksp{1}} \ifdot{\grksp{2}}} \ugsp[2]{123}{2},
\eeq
for some function, $f_{\ifdot{\grksp{1}} \ifdot{\grksp{2}}}$ of \ess{pq}{}{}. 
However,  acting on this with \swqb[1]{3}, implies that both the coefficient of \uth[2]{12}{} and of \uth[2]{23}{} in the resulting expression, vanish independently. In particular, \uth[2]{12}{3} implies that
\beq
\swqb[1]{3}  \lprn \ugsp[1]{123}{1} f_{\ifdot{\grksp{1}} \ifdot{\grksp{2}}} \duinvess{12}{3}{2} \rprn &=& 0.
\eeq
In \sect{partic} we showed that this can only be satisfied when $f_{\ifdot{\grksp{1}} \ifdot{\grksp{2}}} = 0$. One can extend this argument to prove that there are no solutions for $\qsum <  N$ and hence a non-zero $n$-point solution only exists for
\beq
 N \leq \qsum \leq (n-2) N \elab{quenone}.
\eeq 
The upper bound simply follows from the fact that the solution 
must be  composed from 
$(n-2)$ \dgsp[i]{pqr}{}'s, for a given internal symmetry index $i$, 
of which there are $N$ types.

As a result, the general $N=2$ three-point function has to be of the form 
\beq
\left . G_{3}^{(2)} \, \right|_{\qsum = 2} &=&  f(\ess{pq}{}{}) \, \sgsp[1]{123} \, \sgsp[2]{123}, 
\eeq
which is manifestly zero under the action of \swp{1}{1}, \swq[i]{1} and  \swsb[i]{1}. However, using our $N=1$ results, the action of \swqb[1]{1} gives 
\beq
\swqb[1]{1} \left . G_{3}^{(2)} \, \right|_{\qsum = 2} &=&  \frac{\sgsp[1]{123}}{\sess{13}} \, \swqb[1]{1} \lprn \sess{13} \, f(\ess{pq}{}{}) \, \sgsp[2]{123} \rprn  
\eeq
and since \swqb[1]{1} can never be zero on a function of \dgsp[2]{123}{1}, at first sight there  appears to be no solution. 

At this point it helps to consider the action of \swd, which depends on $N$, and hence the $N=2$ solution cannot simply be a product of two $N=1$ solutions, as one might initially expect. Closer inspection reveals that for \swd\ to vanish we require $f$ to be of degree zero in \ess{pq}{}{} and as a result $\sess{13} \, f(\ess{pq}{}{}) \,  \sgsp[2]{123}$ has to be degree zero in \ess{pq}{}{} aswell. Expansion of this term in \dth[2]{p q}{1} gives a series of linearly independent terms whose coefficients can all be expressed as combinations of expressions of the form 
\[
\frac{\esspair{p,p+1}{q,q+1}{1}{2}}{\sess{p,p+1}}.
\]
We know from \sect{uniq} that such functions must vanish under the action of $\sgsp[1]{123} \, \swqb[1]{1}$ at three-points, and thus it is precisely the same nilpotence which prevented us from obtaining unique solutions, that is responsible for the existence of any three-point solutions at all beyond $N=1$. 
Thus, from the $\swqb{1}$  superconformal Ward identity we find no further condition.

As before, we can find the unique form of $f$ by demanding that $\swk{1}{1} G_{3}^{(2)}$ must vanish. It follows from \eqn{l9scwk}, that
\beq
\!\!\!\!\!\!\!\!\!\! \left . \swk{1}{1} \sgsp[1]{123} \sgsp[2]{123} \right|_{\boldq_1 + \boldq_2}  \! \! &=& \! \! \left . \sgsp[1]{123} \lprn \swk[2]{1}{1} \sgsp[2]{123} \rprn \right|_{\boldq_2} +  \left . \lprn \swk[1]{1}{1} \sgsp[1]{123} \rprn \right|_{\boldq_1} \sgsp[2]{123},
\eeq
where, 
\beq
\swk[1]{2}{2} &\equiv& \sum_{p=1}^{n}    \lset s ^{\alpha \dot \beta } s ^{\beta \dot \alpha } \partial_{\alpha \dot \alpha} +s ^{\alpha \dot \beta } \theta ^{\beta (1)} \partial_{\alpha (1) } + q{(4-N)\over N}s^{\beta \dot \beta}  \rset, \, \,\,\,\, \mathrm{etc.} 
\eeq
The $N=1$ results show that 
\beq
\left . \lprn \swk[1]{1}{1} \sgsp[1]{123} \rprn \right|_{\boldq = \boldq_1(N)} &=& 0, \,\,\,\,\,\,\,\,\,\,\,\,    
\boldq_1(N) \,\,\,\, = \,\,\,\, (0, \frac{N}{4-N}, 0),
\eeq
and thus at $N=2$ we find that $\boldq_1(2) = (0,1,0)$, so that $\sgsp[1]{123} \sgsp[2]{123}$ is a solution for 
\beq
\boldq = (0,2,0). \elab{qatfour}
\eeq 
Once again,
\beq
f &=& \sum_{a,b} c_{ab} \lprn \sess{12}\rprn^a \lprn \sess{23}\rprn^b \lprn \sess{13} \rprn^{(-a-b)}, \\
\boldq &=& (0,2,0) + \delta\boldq,
\eeq
from which we deduce that $c_{ab} \neq 0$ only  when 
\beq
a &=& \dq[3]  \,\,\,\,\, = \,\,\,\,\, q_{3}
\nonumber \\
b &=& \dq[1] \,\,\,\,\, = \,\,\,\,\, q_{1}.
\eeq
Therefore, the most general $N=2$ three-point solution is
\beq
G_{3}^{(2)} &=& \lprn \frac{\sess{12}}{\sess{13}} \rprn^{q_3} 
\lprn \frac{\sess{23}}{\sess{13}} \rprn^{q_1} 
\sgsp[1]{123} \sgsp[2]{123}. \elab{l9fullntwosol}
\eeq

\subsection{$N=4$}

In $N=4$ supersymmetry, the $\boldq$ dependence drops out of the Ward Identities because of the factor $(4-N)/ N$ and \swr-symmetry no longer holds. Instead, we can define \qsum\ to be half the degree of \uth[i]{pq}{} and the condition given in \eqn{quenone} still holds.  In the usual way, we write 
\beq
\left . G_{3}^{(4)} \, \right|_{\qsum = 4} &=&  f(\ess{pq}{}{}) \, \sgsp[1]{123} \sgsp[2]{123} \sgsp[3]{123} \sgsp[4]{123}, 
\eeq
and \swd-symmetry implies that 
\beq
f &=&  (\sess{13})^2 \, r(\ess{pq}{}{}),
\eeq
for any $r$ of degree zero in \ess{pq}{}{}.
Written in a different way, the $N=1$ results show that 
\beq
\lbswk[1]{1}{1} \sgsp[1]{123} &=& \ess{2}{1}{1} \sgsp[1]{123},
\eeq
and because $\swk{1}{1} = \lbswk{1}{1}$, when $N=4$, we deduce that
\beq
\lprn \swk{1}{1} - 4 \ess{2}{1}{1}  \rprn \prod_{i=1}^{N} \sgsp[i]{123} &=& 0. 
\eeq
It follows that the general form of the $N=4$ three-point function is 
\beq
G_{3}^{(4)} &=& \lprn \frac{\sess{12}\sess{23}}{\sess{13}} \rprn^2 \prod_{i=1}^{N} \sgsp[i]{123}.
\eeq

\subsection{General Formula}

A concise summary of all our results at three-points is given by the general formula 
\beq
G_{3}^{(N)} &=& \lprn \frac{\sess{12}\sess{23}}{\sess{13}} \rprn^{(N-2)} \lprn \frac{\sess{12}}{\sess{13}} \rprn^{\frac{(4-N)q_3}{N} } 
\lprn \frac{\sess{23}}{\sess{13}} \rprn^{\frac{(4-N)q_1}{N}} \prod_{i=1}^{N} \sgsp[i]{123},
\eeq
with $\qsum = N$.
This formula is also valid for $N=3$.

\newcommand{\gf}{\es{G_{4}}}
\newcommand{\eps}[2]{\es{\epsilon_{\grksp{#1} \grksp{#2}}}}
\newcommand{\epsb}[2]{\es{\epsilon_{\dot{\grksp{#1}} \dot{\grksp{#2}}}}}
\newcommand{\essprod}[4]{\es{ (\jess{#1} .\jess{#2})_{\grksp{#3} \grksp{#4}}}}
\newcommand{\tessprod}[6]{\es{ (\jess{#1} .\jess{#2} .\jess{#3} .\jess{#4})_{\grksp{#5} \grksp{#6}}}}
\newcommand{\essess}[4]{\es{ (\jess{#1} \jess{#2})_{{#3} {#4}}}}
\newcommand{\swks}{\es{K}}
\newcommand{\jth}[1]{\es{\theta_{#1}}}
\newcommand{\jess}[1]{\es{s_{#1}}}
\newcommand{\jessmu}[2]{\es{s_{#1}^{#2}}}
\newcommand{\slst}{\es{\sess{12}, \sess{23},\sess{34}, \sess{13}, \sess{24}, \sess{14}}}

\section{The Four-Point Green's Function at $\qsum = 1$} 

\slab{l9grncalctwo}

The most general four-point function\footnote{In reference \bibnum{pcwfoutnb}, there was some discussion of four-point functions from a different perspective, but the relevant results appear to disagree with those presented here. }, at $\qsum = 1$, satisfying Eqs. \eqnnum{l9scwp}, \eqnnum{l9scwq} and \eqnnum{l9scwsb} is, 

\beq
\gf &=& 
\frac{f}{\sess{12}} \sgsp{123} + 
\frac{g}{\sess{34}} \sgsp{234} + 
\frac{2 h}{\sess{23}} \lprn \ugsp{123}{} \dgsp{234}{} \rprn + 
\frac{4 k}{\sess{12}\sess{23}} \lprn \ugsp{123}{} \ess{12}{}{}  \ess{23}{}{} \dgsp{234}{} \rprn  + \nonumber \\ & &
\frac{4 l}{\sess{12}\sess{23}} \lprn \ugsp{123}{} \ess{23}{}{}  \ess{34}{}{} \dgsp{234}{} \rprn + 
\frac{4 m}{\lprn\sess{23}\rprn^2} \lprn \ugsp{123}{} \ess{12}{}{}  \ess{34}{}{} \dgsp{234}{} \rprn ,
\eeq
where,
\beq
\lprn \ugsp{p}{} \ess{q}{}{}  \ess{r}{}{} \dgsp{t}{} \rprn &\equiv& \ugsp{p}{1} \udess{q}{3}{1}  \duess{r}{3}{2} \dgsp{t}{2}
\eeq
and $f$, $g$, $h$, $k$, $l$, $m$ are all arbitrary functions of the six independent 4-point scalars, 
$\sess{12}$, $\sess{23}$, $\sess{34}$, \essdot{12}{23}, \essdot{23}{34}, \essdot{12}{34}, 
from which all other 4-point scalars can be constructed using the relations given in \app{notapp}. 

We must now impose the rest of the Ward Identities on \gf, beginning with \swd. This implies that all the arbitrary functions $f$, $g$, $h$, $k$, $l$, $m$ are in fact of degree zero, or in other words are functions of five independent ratios of the scalars given above.

To impose \swk{1}{1} is an enormous calculation to perform by hand, and thus we use a computer algebra package, written in Mathematica, especially for the purposes of this calculation. First of all we operate with \swk{1}{1} on \gf, which is relatively straight forward. We shall not go into the details of how that was done in this paper, but we do discuss some of the details of the simplification of the resulting expression in \app{basapp}. 
In particular, this appendix describes how one can define a set of canonical forms in terms of  which all other expressions can be written. In this way, the computer can collect like terms and we can then separate out our results into a sum of linearly independent terms. The coefficients of these terms can then  be set to zero individually, allowing us to restrict the form of the six arbitrary functions using the resulting equations. 

After expanding the expression for $\swk{1}{1} \gf$ in terms of the basis described in \app{basapp}, we can use some of the resulting equations to show that,
\beqa
m &=& 0 \\
h &=& k + l.
\eeqa

After some algebra, it follows that we can rewrite the ansatz for \gf\ in a more symmetric way, as 
\beq
\es{
\gf &=& 
\frac{f'}{\sess{12}} \sgsp{123} + 
\frac{g'}{\sess{34}} \sgsp{234} + 
\frac{k'}{\sess{12}} \sgsp{124} + 
\frac{l'}{\sess{34}} \sgsp{134},
}
\eeq
where $f'$, $g'$, $k'$, $l'$ are degree zero functions of \slst\ (see \eqn{essdotexp}). In addition, the rest of the equations imply that we can  further restrict the form of these functions so that their arbitrariness comes only from their dependence on the four-point cross ratios $u$ and $v$, defined in \eqn{uvdefs}.
We find 
\beqa
\eqal{ f'(\slst) }{=}{ 
Q_0 \lprn \frac{\sess{13}}{\sess{34}} \rprn^3 \lprn \frac{\sess{24}}{\sess{14}} \rprn^2 f''(u,v)}
\eqal{ g'(\slst) }{=}{
Q_0 \lprn \frac{\sess{13}}{\sess{14}} \rprn^2 g''(u,v)}
\eqal{ k'(\slst) }{=}{
Q_0 \lprn \frac{\sess{13}}{\sess{34}} \rprn^3 \lprn \frac{\sess{24}}{\sess{14}} \rprn^2 k''(u,v)}
\leqal{ l'(\slst) }{=}{ 
Q_0 \lprn \frac{\sess{13}}{\sess{14}} \rprn^2 l''(u,v)},
\eeqa
where
\beq
Q_0 &=& \frac{
\lprn \sess{34} \rprn^{3\,\lprn q_{1} + q_{2} \rprn} 
\lprn \sess{14} \rprn^{3\,\lprn q_{2} + q_{3} \rprn} 
}{
\lprn \sess{13} \rprn^{3\,\lprn q_{1} + q_{2} + q_{3} \rprn} 
\lprn \sess{24} \rprn^{3\, q_{2}}
}.
\eeq
 
Further restrictions can then be found by imposing \swqb{1} on \gf, in addition to the above restrictions from \swks.
The result is the following set of equations for the undetermined functions of $u,v$, 

\beq
& & 
\ba{rcl}
\eqal{\p_{v}f }{=}{ 
   {{f\,\left( -1 + 3\,q_{1} + 3\,q_{4} \right) }\over {v}} + }
\eqal{}{}{ 
     {{-6\,{u^2}\,g\,q_{1} + 3\,{u^2}\,z\,l\,q_{2} - 
         3\,y\,k\,q_{3} + 
         3\,\left( 1 + u - v \right) \,f\,q_{4}}\over 
       {w} }
}
\leqal{
 \p_{u}f }{=}{ 
   {{3\,f\,\left( 1 - q_{1} - q_{2} \right) }\over {u}} - 
     {{3\,\lprn u\,y\,g\,q_{1} - u\,v\,x\,l\,q_{2} + 
         2\,v\,k\,q_{3} - z\,f\,q_{4} \rprn}\over {w}}
}
\elab{qf}
\ea  
\\ \nonumber \\
\nonumber \\
\nonumber \\
& & \ba{rcl}
\eqal{
 \p_{v}g }{=}{  
   {{g\,\left( 2 - 3\,q_{2} - 3\,q_{3} \right) }\over {v}} + 
     {{3\,u\,x\,g\,q_{1} - 3\,u\,y\,l\,q_{2} + 3\,z\,k\,q_{3} - 
         6\,f\,q_{4}}\over {u\,w}}
}
\leqal{   
 \p_{u}g }{=}{  
   {{3\,{u^2}\,z\,g\,q_{1} - 6\,{u^2}\,v\,l\,q_{2} + 
        3\,v\,x\,k\,q_{3} - 3\,y\,f\,q_{4}}\over {u^{2}\,w}}
}
\elab{qg}
\ea 
\\ \nonumber \\
\nonumber \\
\nonumber \\
& & \ba{rcl}
\eqal{
{\p_{v}k} }{=}{  
   {{{3\,{u^2}\,z\,g\,q_{1} - 6\,{u^2}\,v\,l\,q_{2} + 
        3\,v\,x\,k\,q_{3} - 3\,y\,f\,q_{4}}\over {v\,w}}}
}
\leqal{
 {\p_{u}k} }{=}{  
   {{{3\,k\,\left( 1 - q_{1} - q_{2} \right) }\over u} + 
     {{3\,u\,x\,g\,q_{1} - 3\,u\,y\,l\,q_{2} + 3\,z\,k\,q_{3} - 
         6\,f\,q_{4}}\over w}}
}
\elab{qk}
\ea 
\\ \nonumber \\
\nonumber \\
\nonumber \\
& & \ba{rcl}
\eqal{
 {\p_{v}l} }{=}{  
   {{{-3\,u\,y\,g\,q_{1} + 3\,u\,v\,x\,l\,q_{2} - 
        6\,v\,k\,q_{3} + 3\,z\,f\,q_{4}}\over {u\,v\,w}}}
}
\leqal{  
 {\p_{u}l} }{=}{  
   {{{-6\,{u^2}\,g\,q_{1} + 3\,{u^2}\,z\,l\,q_{2} - 
        3\,y\,k\,q_{3} + 3\,x\,f\,q_{4}}\over {{u^2}\,w}}}
}
\elab{ql}
\ea
\eeq

where, for clarity,  we have dropped the double primes for the rest of the discussion and 
\beqa
x &=& 1 + u - v  \\
y &=& 1 - u - v  \\
z &=& 1 - u + v  \\
w &=& 1 - 2\,u + {u^2} - 2\,v - 2\,u\,v + {v^2}.
\eeqa

It is immediately apparent that by specifying four constants, which are the values of $f$, $g$, $k$ and $l$ at some point $u_0, v_0$, we can determine all first derivatives using Equations \eqnnum{qf} - \eqnnum{ql}. By differentiation we can determine all second derivatives in terms of known lower derivatives at $u_0, v_0$ and thus we can determine all higher derivatives at this point by repeating this process indefinitely. Consequently we can construct the solution around $u_0, v_0$ as an infinite Taylor expansion in $u,v$, and thus the solution is uniquely specified by these four constants. 

Of course one could instead try to solve the above equations  by imposing integrability relations, such as 
\beq
\frac{\partial^{2}}{\partial u \partial v} f &=& \frac{\partial^{2}}{\partial v \partial u} f .
\eeq
By calculating $\frac{\partial^{2}}{\partial v \partial u} f $ in two different ways from \eqnnum{qf} and equating these expressions, we ought to find a relationship between $f$, $g$, $k$, $l$ and their first derivatives. Substituting for the first derivatives using  Equations \eqnnum{qf} - \eqnnum{ql}, we should get a relationship between $f$, $k$, $g$ and $l$ alone. However, we find in all cases that this is simply the trivial statement that $0=0$. This implies that the equations are integrable in the given form and thus in order to solve them one must specify the values of $f$, $g$, $k$ and $l$ at some initial point, as above. In other words, the integrability relations do not yield further relationships which could be used to reduce the number of constants which have to be specified to obtain a solution.

An alternative approach is to consider Equations \eqnnum{ql}, and rewrite them to make  $k$ and $g$ the subjects, as in  
\beqa
\eqal{\!\!\!\!\!\!\! {g\,q_{1}} }{=}{ 
     {{{u\,v\,\left( -1 + u + v \right) \,\p_{v}l_{2} + 2\,{u^2}\,v\,\p_{u}l_{2} + 
   3\,u\,v\,l_{2}\,q_{2} + 3\,f_{2}\,q_{4}}\over {3\,u}}} }
\leqal{\!\!\!\!\!\!\!    {k\,q_{3}} }{=}{
     {{{2\,{u^2}\,v\,\p_{v}l_{2} + {u^2}\,\left( -1 + u + v \right) \,\p_{u}l_{2} + 
   3\,{u^2}\,l_{2}\,q_{2} + 3\,f_{2}\,q_{4}}
         \over 3}}
}
\eeqa
We then substitute these expressions into Eqs. \eqnnum{qf} to eliminate $k$ and $g$ completely, giving 
\beqa
\eqal{ {\p_{u}f} }{=}{ 
     {{{{u^2}\,v\,\p_{v}l + 
          \left( 3 - 3\,q_{1} - 3\,q_{2} \right)\, f }\over u}} }
\leqal{
 {\p_{v}f} }{=}{  
     {{{{u^2}\,v\,\p_{u}l + 
          \left( -1 + 3\,q_{1} + 3\,q_{4} \right)\, f }\over v}}.
}
 \elab{qfprime}
\eeqa
Note that at no time do we divide by any function of $q_{i}$ as without knowing \boldq,  we cannot be sure that such a function is non-zero. 
Differentiating the first of Eqs. \eqnnum{qfprime} with respect to $v$ and the second with respect to $u$, we use the integrability condition 
\beq
\frac{\partial^{2}}{\partial u \partial v} f &=& \frac{\partial^{2}}{\partial v \partial u} f 
\eeq
to combine the two equations. We then use Eqs. \eqnnum{qfprime} to remove the derivatives of $f$ from the resulting expression, giving an equation in $l$ alone, 
\beq
 \!\!\!\! v\,\p_{{v^2}}l - u\,\p_{{u^2}}l + 
   \left( 1 - 3q_{1} - 3q_{2} \right) \, \p_{u}l + 
   \left( 2 - 3q_{1} - 3q_{4} \right) \, \p_{v}l &=& 0, 
\elab{lone}
\eeq
where 
\beq
\p_{u^n v^m} l &\equiv& \frac{\partial^{n+m}}{\partial u^n \partial v^m}.
\eeq

To do this, we only used Equations \eqnnum{ql} and \eqnnum{qf}, and thus we could repeat the process using Equations \eqnnum{ql} and \eqnnum{qg} or Equations \eqnnum{ql} and \eqnnum{qk} to obtain other equations in $l(u,v)$ alone. Of the three second order linear partial differential equations which result, only two are linearly independent. The other can be written as 
\beqa
 \left( -1 + v \right)\, v \,\p_{{v^2}}l + 2\, u v \,\p_{u v}l + 
   {u^2}\,\p_{{u^2}}l + 3\, u \,
    \left( q_{1} + 2 q_{2} + q_{3} \right) \,\p_{u}l +  & & \\ 
   3 \, q_{2}\,\left( 2 - 3 q_{4} \right)\, l  + 
   \left( 1 - 3\,q_{2} - 3\,q_{3} + 3\,v\,\left( 1 + q_{2} - q_{4} \right) \right)\, \p_{v}l &=& 0
\elab{ltwo}
\eeqa

To investigate these simultaneous equations we proceed as follows.\footnote{We wish to thank Thomas Wolf and Allan Wittkopf for their help with the following argument.}
Once again, we continue to try to impose higher order integrability relations, such as 
\beq
\frac{\partial^{3} l}{\partial u^2 \partial v} &=& \frac{\partial^{3} l}{\partial v \partial u^2} \,\,\,\,\,\, etc, 
\eeq
by differentiating the above pde's for $l$. We make $\p_{u^2} l$ the subject of \eqn{lone} and $\p_{uv} l$ the subject of \eqn{ltwo}. Differentiating \eqn{lone} with respect to $v$ and \eqn{ltwo} with respect to $u$ and eliminating $\p_{u^2 v} l$, we obtain three differential equations for \[ \p_{uv} l, \,\,\,\, \p_{u^2} l, \,\,\,\,  \p_{v^3} l \] (in terms of only $l, \p_{u} l, \p_{v} l, \p_{v^2} l$), from which all other higher derivatives can be obtained. That is to say, we have a system of pde's whose integrability conditions are simply identities which follow as a consequence of these three equations alone and thus impose no further restrictions on $l(u,v)$. This corresponds to the statement above where the  complete solution is determined by four independent arbitrary constants. 

In this approach one can construct complete solutions given a function $l$ which satisfies the given differential equations. These solutions are not in the form of Taylor expansions and,  as an illustration, we have explicitly constructed two distinct solutions for a given \boldq. For example, if we take 
\beq
\boldq &=& (-1/3, 0, q_3, 4/3 - q_3)
\eeq
we find that 
\beqa
k &=& \frac{4 - 3q_3}{3 q_3} u^4 v^{(2-3q_3)}  \\
g &=& (3q_3 - 4) u^4 v^{(2-3q_3)}  \\ 
f &=& u^4 v^{(2-3q_3)} \\
l &=& \mathrm{constant,}
\eeqa
is a solution, and so is
\beqa
k &=& (2 + 2 u - 3 q_3 u - 2 v) (2 - 3q_3) v^{(2-3q_3)} \\
f &=& - \frac{1}{2} (2 - 3q_3) u^2 v^{(2-3q_3)} \\
g &=& \frac{1}{2} (2 - 3q_3) u^2 v^{(2-3q_3)}  \\ 
l &=& v^{(2 - 3q_3)}.
\eeqa
Clearly these two solutions have the same \boldq, but they differ in the values of the four constants which determine the particular form of the solution. One can construct similar examples for different choices of \boldq.

\section{Conclusions}

In this paper we have found the most general three-point Green's function 
for $N= 1,2,3,4$ supersymmetry which is composed of 
chiral superfields of a given chirality.  We have also shown that although there exist no chiral superconformal invariants, \bibnum{l9scfpap8} the Green's functions of chiral superfields are not uniquely specified above three-points when the the $R$-charge, \qsum, is greater than $N$. This result relies crucially on the nilpotent character of such Green's functions.  However, for the particular case $\qsum = 1, N = 1$ we have shown that the solutions are unique  up to the specification of four constants of integration. We have given two equivalent formulations of the solution which should enable one to explicitly construct the solution in any particular case. 

The results of our investigations seem to suggest that our findings should generalise to extended supersymmetry, where we expect to find a unique solution at four-points in the case where $\qsum = N$. Furthermore, it is tempting to suggest that all higher-point functions may also be uniquely determined in the case where $\qsum = N$. At the moment, however, these two statements remain conjectures based upon our explicit results for the $N=1$ four-point solution at $\qsum = N$ and for the three-point solutions for $N \geq 1$.

\section{Acknowledgements}

We wish to thank Paul Howe for useful discussions. We also wish to thank Alain Moussiaux at CONVODE and Edgardo S. Cheb-Terrab for advice on the solution of partial differential equations. In particular, we wish to thank Thomas Wolf and Allan Wittkopf for help with their pde solving packages CRACK and RIF, which eventually solved the system of pde's discussed in the paper. AP wishes to thank D.R.T. Jones for discussions and PPARC for a research fellowship.

\section{Note Added}

After this work had been completed a preprint \bibnum{parktwo} appeared on the hep-th archive which also discusses Green's functions in superconformal field theories.

\myappendix
\appsection{A Linearly Independent Basis}

\alab{basapp}

In this paper, we need to define a set of linearly independent terms in terms which any relevant expression up to four points can be expanded. This basis of terms can be used to define an explicit expression for a Green's function which is written as a sum of such terms with  arbitrary scalar functions as coefficients. Each Ward Identity then gives us a single constraint on each Green's function, but by expansion in an appropriate basis of terms one can deduce several differential equations which must be independently satisfied by the unknown functions in the expansion of the Green's function. Therefore it is critical that we are able to define a suitable collection of bases to justify  all  of the calculations performed in the paper. In this Appendix we discuss the main details of how this was done. 

The definition of linear independence is that if a linear combination of a set of linearly independent terms vanishes, then all coefficients must vanish independently. Hence, to prove that a set of terms are linearly independent we need to consider general expressions at up to four points which are known to vanish. The Ward Identities also provide us with a set of expressions which are known to vanish, and thus problem of the determination of a basis for expansion of a generalised Green's function and the problem of separation of a Ward Identity into several independent identities for the unknown coefficients in the Green's function can be solved in the same way. This is done as follows. 
 
Imagine that we have an expression in $\uth{p}{1}$ and \ess{p}{1}{1} which vanishes, such as a Ward Identity. We need to simplify this expression so that  it is written in terms of a linearly independent set of terms with some known coefficients. 
To begin with we observe that one can always consider terms containing different $\theta$ variables to be linearly independent. For example, the equation
\beq
a \jth{1}^2 + b \jth{2}^2 + c \jth{1} \jth{2} &=& 0, 
\eeq
where $a$, $b$, $c$ are independent of $\jth{i}$, implies that $a = b = c = 0$. This is easily shown by multiplying by $\jth{1}^2$, to show that $b=0$, and then continuing in the obvious way with $\jth{1}\jth{2}$ and $\jth{1}^2$. 

It follows that we can immediately separate out a given vanishing expression  into coefficients of distinct combinations of $\uth{p}{1}$ and equate these coefficients to zero. (In all our expressions we ensure, for simplicity,  that there are no free indices by contracting them with the arbitrary vector $k^{grksp{1}\dot{grksp{1}}}$ or sometimes an inert $\theta^{\alpha}$ variable.) 
We write each term in the form
\beq 
\mth{q_1}{2} \ldots \mth{q_m}{2} \mth{p_1}{\alpha} \mth{p_2}{\beta} \ldots \mth{p_n}{\gamma} h_{\alpha\beta \ldots \gamma} &=& 0
\eeq
and define $\alpha < \beta < \ldots < \gamma$, given that $p_1 < p_2 < \ldots < p_n$. 
The coefficients, $h_{\alpha\beta \ldots \gamma}$, are in general functions of \ess{p}{1}{1} and may be written in many different but equivalent forms.  

To give a simple example, consider the relation,
\beq
\eps{1}{2} \eps{3}{4} - \eps{1}{3} \eps{2}{4} + \eps{2}{3} \eps{1}{4} &=& 0.
\eeq
Clearly, we must not allow one of these terms on the left to be present in  the expansion of $h_{a_1 a_2 \ldots a_n}$, as it is linearly dependent on the other two. We therefore impose the rule that, in the canonical (i.e. simplified) form of an expression,
\beq
\eps{1}{2} \eps{3}{4} &\rightarrow& \eps{1}{3} \eps{2}{4} - \eps{2}{3} \eps{1}{4},
\eeq
if $\grksp{1} < \grksp{3}$ and  $\grksp{2} > \grksp{4} > \grksp{3}$. 

Note that we are also assuming that the antisymmetry of $\eps{1}{2}$ has already been used to order the indices uniquely in each $\eps{}{}$ tensor on the left, so that $\grksp{1} < \grksp{2}$ and  $\grksp{3} < \grksp{4}$. 
Similarly, 
we can use the relations in \app{notapp} to show that any product of \ess{p}{1}{1} can be reduced to a sum of bilinear products of the form  \essprod{p}{q}{1}{2} and scalars. Such terms can be written such that $p < q$ and $\grksp{1} < \grksp{2}$, up to terms involving only scalars and \eps{1}{2} (see \app{notapp}). These latter terms are lower order in \essprod{p}{q}{1}{2}  and can  subsequently be  further ordered without producing any further higher order terms. (In this respect, since the number of free indices in each term is always  the same, we define lower order terms as those containing more \eps{1}{2} terms and fewer \essprod{p}{q}{1}{2} and vice versa for higher order terms.) Given this,  we impose the rule,
\beq
\eps{3}{4} \essprod{p}{q}{1}{2} &\rightarrow& \eps{3}{1} \essprod{p}{q}{4}{2} +  \eps{1}{4} \essprod{p}{q}{3}{2},  \elab{rlthr}
\eeq 
whenever $\grksp{1} < \grksp{3} < \grksp{4}$ (assuming $\grksp{3} < \grksp{4}$ and $\grksp{1} < \grksp{2}$),
to define a canonical form for products of \essprod{p}{q}{1}{2} and $\eps{1}{2}$.
Then, we use the relations 
\beq
\essprod{p}{q}{1}{2}
\essprod{r}{t}{3}{4} &\rightarrow&
\essprod{p}{q}{1}{3}
\essprod{r}{t}{2}{4}
- \tessprod{p}{q}{r}{t}{1}{4} \eps{2}{3},
\eeq
where $p < q$, $r < t$, $c < b$ and $\essprod{p}{q}{1}{2} \neq \essprod{r}{t}{1}{2}$, and
\beq
\essprod{p}{q}{1}{2}
\essprod{p}{q}{3}{4} &\rightarrow&
\essprod{p}{q}{1}{3}
\essprod{p}{q}{2}{4}
- \tessprod{p}{q}{p}{q}{1}{4} \eps{2}{3},
\eeq
where $\grksp{1} < \grksp{3} < \grksp{2}$, to show that any product of  \essprod{p}{q}{1}{2} tensors can have its indices totally ordered. Once again, the extra terms on the right hand side are of lower order in \essprod{p}{q}{1}{2} and can therefore be further ordered but without generating terms of higher order than themselves (see \eqn{essfourexp}). 

 Next we must deal with the internal dotted indices in the \essprod{p}{q}{1}{2} terms. One can use the relation 
\beq
\epsb{1}{2} \epsb{3}{4} &=& \epsb{1}{3} \epsb{2}{4} - \epsb{2}{3} \epsb{1}{4},
\eeq
to define the rule 
\beq
\essprod{p}{q}{1}{2}
\essprod{r}{t}{3}{4} &\rightarrow&
\essprod{p}{r}{1}{3}
\essprod{q}{t}{2}{4}
- \essprod{p}{t}{1}{4}
\essprod{q}{r}{2}{3},
\eeq
whenever $t < q$ and $r < p$, assuming $p < q$ and $r < t$. This is compatible with \eqn{rlthr}.

The above rules apply to all terms, even those with more than four free indices, but they do not necessarily define a basis as we have not shown that the set of allowed terms which remain are  linearly independent. Clearly the remaining terms span the space, as we can write down all possible terms which could arise and then use the above rules to reduce them to a much smaller set of terms, which therefore manifestly span the space. 
It remains to find a linearly independent subset of the resulting set of terms. However,  it is not always immediately obvious whether a given set of terms is linearly independent or not. Some of the linear dependencies which exist amongst the canonical terms above are rather complicated. For example, there are two ways to simplify the following trace 
\beq
(\ess{1}{}{} . \ess{2}{}{} . \ess{3}{}{} . \ess{4}{}{} . \ess{4}{}{} . \ess{3}{}{} . \ess{2}{}{} . \ess{1}{}{}).
\eeq
An obvious method of simplification is to use \eqn{l9spsqu} repeatedly to show that 
\beq
(\ess{1}{}{} . \ess{2}{}{} . \ess{3}{}{} . \ess{4}{}{} . \ess{4}{}{} . \ess{3}{}{} . \ess{2}{}{} . \ess{1}{}{}) &=& -\frac{1}{8} \sess{1} \sess{2} \sess{3} \sess{4}. \elab{scalld}
\eeq
Alternatively, one might first use \eqn{essfourexp} on the first four terms and then separately on the last four terms before multiplying out the result. It is clear by inspection that this has to give an expression in $[s_{1}.s_{2}.s_{3}.s_{4}]^2$, which is not manifestly the same as  the right-hand side of \eqn{scalld}, but is equal to it by construction. Consequently, there is a non-trivial relationship between a set of scalar quantities, and we can use this in our calculations to eliminate $[s_{1}.s_{2}.s_{3}.s_{4}]^2$, for example. 

In a similar way, one can construct relationships between the canonical terms defined above which allow us to eliminate all terms of the form 
\beq
[s_{1}.s_{2}.s_{3}.s_{4}] \essprod{1}{2}{1}{2}
\eeq 
from our basis. (We note that this can be done in all the cases we consider in this paper as we can ensure that we have only four independent four vectors, namely $k$, \ess{12}{}{}, \ess{23}{}{} and \ess{34}{}{}. In calculations involving \swk{}{} we must also include \ess{0}{}{}, but one can easily show that $\swk{}{} G$ is independent of \ess{0}{}{} as long as $\swd G = 0$, which we can always choose to be true where necessary.)

Even given the above results, finding all of the linear dependencies which can arise between the terms which remain is a difficult task and since the proof of linear independence is vital to our approach we set out to prove the validity of any suggested basis as follows. 
Beginning with a set of terms, $v^{i}_{abc\ldots}$, with some fixed number of free indices, $abc\ldots$, which span the space and which have all known linear dependencies removed, in the way described above, we can write down the most general expression $T_{abc\ldots}$ with this number of free indices as an expansion in $v^{i}_{abc\ldots}$ with arbitrary scalar coefficients, $a_i$ as follows
\beq
T_{abc\ldots}  &=& \sum_{i} v^i_{abc\ldots} a_i.
\eeq
Note that the $a_i$ are scalar functions of \ess{12}{}{}, \ess{23}{}{} and \ess{34}{}{} only and any dependence of $T$ on $k$, which always occurs at most linearly in $T$ for our calculations, must be carried explicitly by the basis terms, $v^i$. Therefore we need a $k$-dependent basis for Ward Identitity expansions and a $k$-independent basis for Green's function expansions. A $k$-independent basis can be easily deduced from the corresponding $k$-dependent basis, i.e.\ the one with the same number of free indices on $T$. 

We are concerned with the conditions satisfied by the functions $a_i$, given that $T = 0$. By contracting each of the $v^i$ with $T$ we can build up a set of equations $v^{iabc\ldots}T_{abc\ldots} = 0$. Furthermore, by eliminating $(k.s_{12}.s_{23}.s_{34})^2$ as described above, following \eqn{scalld}, we can equate the coefficients of $(ks_{12})$,  $(ks_{23})$,  $(ks_{34})$  and $(k.s_{12}.s_{23}.s_{34})$ to zero independently and hence generate several equations in each case. This step can be rigorously justified by choosing several independent values for the arbitray vector $k$ and solving the resulting equations in each case, but note that it is clearly invalid unless we eliminate $(k.s_{12}.s_{23}.s_{34})^2$ first.

The resulting set of equations can be written as
\beq
M_{ij} a_j &=& 0, \elab{basmat}
\eeq
for some matrix $M$ dependent only on \ess{12}{}{}, \ess{23}{}{} and \ess{34}{}{}. We can then attempt to put $M$ into upper triangular form using row and column operations and in the process simply removing any rows which are all zero as these correspond to trivial conditions on the $a_i$. However, we must take great care that we do not divide by zero inadvertently during this process as it may be that we divide by a scalar expression which is not manifestly zero, but which vanishes upon expansion into the  components of the four-vectors of which it is constructed. (The discussion following \eqn{scalld} illustrates such a null scalar expression composed of four four-vectors. It is unlikely that such an expression can be constructed from only three four-vectors, but we were in any case careful to explicitly avoid this error. Such considerations will clearly be relevant for studies of five-point functions and above.) 

If the matrix, $M$,  can be reduced to upper triangular form, its determinant can be  easily calculated as the product of the diagonal elements and by expansion in components can be shown not to vanish. In this case, the $v^i_{abc\ldots}$ form a basis as we have demonstrated that when $T = 0$, $a_i = 0$. 

If the matrix, $M$, cannot be reduced to upper triangular form we can use \eqn{basmat} to eliminate a maximal subset of the $a_i$ from $T$. Since $T$ vanishes and the remaining $a_i$ are arbitrary, it follows that the coefficient of each   of the remaining $a_i$ is a linear dependency amongst the original $v^i$ and can be used to eliminate some of the $v^i$ to define an improved set of terms ${v'}^{i}$ which are, by construction, manifestly linearly independent and span the space, i.e.\ a basis. 

In summary, this process allows us to both define a basis in terms of which to write a general Green's function before imposing the Ward Identities or alternatively to define a basis in terms of which we can expand a particular Ward Identity, and also it generates any extra simplification rules necessary to reduce a particular expression to the required canonical form, so that we are justified in  equating the coefficients of the resulting terms to zero. 
It is these sort of calculations which validate the various arguments given in the rest of the paper.

\remark{

As a result of these rules we might expect to be able to reduce all terms to a linearly independent basis set of terms of the form of products of $\eps{1}{2}$ and , \essprod{p}{q}{1}{2}, whose coefficients we can extract, and which will be scalar functions, equal to zero. However, there is a subtlety when more than four 4-vector terms are involved in a scalar expression. To see this,  consider the following identity,
\beq
\eta^{\mu [\nu} \epsilon^{\alpha \beta \gamma \delta]} &=& 0
\eeq
 which is true simply because there is no tensor which is totally antisymmetric on five indices, in four dimensions. Soaking up the 4-vector indices by multiplying by different $s_{1}^{\mu}$ etc, one can construct all sorts of non-trivial relationships between fuctions of \ess{p}{1}{1}, which ruin linear independence of the canonical forms defined above. (See \app{notapp} for details of how $\epsilon^{\alpha \beta \gamma \delta}$ etc. arises from terms in \ess{p}{1}{1}). However, if we have only four distinct 4-vectors, then we can easily see that any such relationship must become a trivial identity, because two of the five 4-vectors must be identical and we have therefore symmetrised on at least two of the five antisymmetric indices already, giving zero on both sides of the equation. 
  
Hence, we deduce that we need not consider these relationships if we have less than five independent 4-vectors. Unfortunately, at four points the expression for $\swks \, \gf$ is dependent upon five distinct 4-vectors, and there is potentially a problem. 

There are three distinct 4-vectors, \jess{12}, \jess{23}, \jess{34} already present in \gf. 

?? austin what i am getting at is that there is a four four vector 
which is $\epsilon$ contracted on three of its indices with the 
three $s_{ij}$ . i am not sure if this makes a difference but we forgot it in an earlier paper and got a worng result. if one expands $T_{\alpha \dot \beta}$ 
it must be there.

The other two are \jess{0}, which arises because \swks\ and \swp{1}{1} do not commute, and the dummy variable $\kay{1}{1}$, which we use  to soak up the indices on \swk{1}{1} to give
\beq
\swks &=& \swk{1}{1} \kay{1}{1}.
\eeq
The advantage of this is that we don't need to consider free indices. $\kay{1}{1}$ is of course arbitrary and can be removed once the expression has been simplified. It is needed to make the simplification process described above work consistently, and cannot be ignored as the definitions of the canonical forms depend on the absence of free indices in the overall expression. 

It turns out that when we calculate $\swks\, \gf$, and simultaneously impose $\swd\, \gf = 0$, the dependence on \jess{0} disappears. This is of course obvious from the superconformal algebra, where the commutator of \swp{1}{1} and \swks\ is \swd. Thus, when $\swd\, \gf = 0$ and $\swp{1}{1}\, \gf = 0$, it follows that $\swp{1}{1}\, \swks\, \gf = 0$ and \jess{0} is not present. What is not obvious is that this should be manifestly the case in the computer output without the need for expressions generated by symmetrising on five indices, as above. (To give a counter example, at five points the computer cannot prove the obvious result that \swsb{1} vanishes on a function of \ugsp{pqr}{1}, unless one imposes such relations.) In fact, a careful study of the action of \swks\ on \gf\ shows how this comes about and one can predict it without doing the full calculation. Thus, we can state that $\swks\, \gf$ is dependent on only four distinct  4-vectors and we need not impose any of the relations derived by anti-symmetrising expressions on five independent indices. 

}

\appsection{Notation and Useful Identities}

\alab{notapp}

We present below an explanation of our notation and some relations which we found particularly useful.

\subsection{Notation}

\beq
\essprod{p}{q}{1}{2} &\equiv& \duess{p}{1}{3} \ddess{q}{2}{3} \\
\essdot{p}{q} &\equiv& \ess{p}{1}{2} \ddess{q}{1}{2} \\
\tessprod{p}{q}{r}{t}{1}{2} &\equiv&  \duess{p}{1}{3} \udess{q}{4}{3} \duess{r}{4}{5} \ddess{t}{2}{5} \\
(s_{p}. s_{q} . s_{r} . s_{t}) &\equiv&  \ess{p}{2}{3} \udess{q}{4}{3} \duess{r}{4}{5} \ddess{t}{2}{5} \\
\ddess{p}{1}{1} &\equiv& s_{\mu} \sigma^{\mu}_{\grksp{1}\ifdot{\grksp{1}}} \\
s_{p\ifdot{\grksp{1}}\grksp{1}} &\equiv& s_{\mu} \bar{\sigma}^{\mu}_{\ifdot{\grksp{1}}\grksp{1}} 
\eeq
\beq
\lprn \ugsp{pqr}{} \dgsp{nlm}{} \rprn &\equiv& \ugsp{pqr}{1} \dgsp{nlm}{1}
\\
\ugsp[i]{pqr}{1} &\equiv& \uth[i]{pq}{2} \duinvess{pq}{2}{1} - \uth[i]{qr}{2} \duinvess{qr}{2}{1} 
\\
\lprn \ugsp{p}{} \ess{q}{}{}  \ess{r}{}{} \dgsp{t}{} \rprn &\equiv& \ugsp{p}{1} \udess{q}{3}{1}  \duess{r}{3}{2} \dgsp{t}{2}
\eeq

\subsection{Useful Relations}

\beq
\ugsp{pqr}{1} &=& - \ugsp{rqp}{1} \\
\ugsp{pqr}{1} &=& \ugsp{prq}{2} \udess{pr}{3}{2} \duinvess{pq}{3}{1} \\
\ugsp{pqr}{1} &\propto& \ugsp{\langle pqr \rangle}{1},
\eeq
where $\langle pqr \rangle$ is the ordered form of $pqr$ ( e.g. $\langle 312 \rangle = 123$). 

\beq
\essprod{p}{q}{1}{2} &=& \lprn \eta^{\mu\nu} \eps{1}{2} + \sigma^{\mu\nu}_{\grksp{1}\grksp{2}}  \rprn s_{p\mu} s_{q\nu} \\
\ess{p}{1}{1} &=& s_{p}^{\ifdot{\grksp{1}}\grksp{1}}
\eeq
  
\beqa
\eqal{
\duess{12}{1}{1} \ddess{12}{2}{1} }{=}{ - \frac{1}{2} \sess{12} \scfeps{1}{2} 
}
\leqal{
\udess{12}{1}{1} \ddess{12}{1}{2} }{=}{ - \frac{1}{2} \sess{12} \scfepsb{1}{2}  \elab{l9spsqu},
}
\eeqa
\beq
\duess{12}{1}{2} \ddess{23}{3}{2} \udess{12}{3}{1} &=& \essdot{12}{23} \ddess{12}{1}{1} - \frac{1}{2} \sess{12} \ddess{23}{1}{1}. \elab{l9spswap}
\eeq
\beq
{(s_{2} . s_{1})}_{\h_{2}\h_{1}} &=& - {(s_{1} . s_{2})}_{\h_{1}\h_{2}} 
\eeq
\beq
{(s_{1} . s_{2})}_{\h_{2}\h_{1}} &=& {(s_{1} . s_{2})}_{\h_{1}\h_{2}} + (s_{1}s_{2})\, \e_{\h_{1}\h_{2}}
\eeq

\beq
 2 \lprn s_{1}.s_{2}.s_{3}.s_{4} \rprn_{\h_{1}\h_{2}} &=& 
- {(s_{3} . s_{4})}_{\h_{1}\h_{2}}\,(s_{1}s_{2})   + 
      {(s_{2} . s_{4})}_{\h_{1}\h_{2}}\,(s_{1}s_{3}) - \nonumber \\ & &
      {(s_{2} . s_{3})}_{\h_{1}\h_{2}}\,(s_{1}s_{4}) - \nonumber \\ &  &
      \left( {(s_{1} . s_{4})}_{\h_{1}\h_{2}} + 
         \e_{\h_{1}\h_{2}}\,(s_{1}s_{4}) \right) \,(s_{2}s_{3}) + \nonumber \\ & & 
      \left( {(s_{1} . s_{3})}_{\h_{1}\h_{2}} + 
         \e_{\h_{1}\h_{2}}\,(s_{1}s_{3}) \right) \,(s_{2}s_{4}) - \nonumber \\ &  &
      \left( {(s_{1} . s_{2})}_{\h_{1}\h_{2}} + 
         \e_{\h_{1}\h_{2}}\,(s_{1}s_{2}) \right) \,(s_{3}s_{4}) - \nonumber \\ & & 
      \e_{\h_{1}\h_{2}}\,(s_{1}.s_{2}.s_{3}.s_{4}) \elab{essfourexp}
\eeq
\beqa
\eqal{
\tessprod{p}{q}{r}{t}{1}{2} }{=}{  \frac{1}{2} \essdot{p}{r} \essprod{q}{t}{1}{2} - \frac{1}{2} \essdot{p}{q} \essprod{r}{t}{1}{2} 
}
\leqal{}{}{ \mbox{} - \frac{1}{2} \essdot{q}{r} \essprod{p}{t}{1}{2} + i \epsilon^{\mu\nu\rho\kappa} s_{p\mu} s_{q\nu} s_{r\rho} {\sigma_{\kappa\grksp{1}}}^{\ifdot{\grksp{1}}} \ess{t}{2}{1}
}
\eeqa

\newlength{\ifempa}
\newlength{\ifempb}
\settowidth{\ifempb}{\mbox{$a()$}}
\newcommand{\ifemp}[1]{\settowidth{\ifempa}{\mbox{\ensuremath{a(#1)}}}
\ifthenelse{\lengthtest{\ifempa = \ifempb}}{}{(#1)}
}
\newcommand{\inp}{in preparation}
\newcommand{\tbp}{to be published}
\newcommand{\npb}[3]{Nucl.\ Phys.\ {\bfseries B#1} (#2) #3}
\newcommand{\pr}[3]{Phys.\ Rev.\ {\bfseries #1} \ifemp{#2} #3}
\newcommand{\prl}[3]{Phys.\ Rev.\ Lett.\ {\bfseries #1} \ifemp{#2} #3}
\newcommand{\cmp}[3]{Commun.\ Math.\ Phys.\ {\bfseries #1} \ifemp{#2} #3}
\newcommand{\pl}[3]{Phys.\ Lett.\ {\bfseries #1} \ifemp{#2} #3}
\newcommand{\plb}[3]{Phys.\ Lett.\ {\bfseries #1B} \ifemp{#2} #3}
\newcommand{\mpla}[3]{Mod.\ Phys.\ Lett.\ {\bfseries A#1} \ifemp{#2} #3}
\newcommand{\jetpl}[3]{J.\ E.\ T.\ P.\ Lett.\ {\bfseries #1} \ifemp{#2} #3}
\newcommand{\ijmp}[3]{Int.\ Journal.\ Mod.\ Phys.\  {\bfseries #1} \ifemp{#2} #3}
\newcommand{\jpa}[3]{J.\ Phys.\ {\bfseries A#1} \ifemp{#2} #3}
\newcommand{\tmf}[3]{Theor.\ Mat.\ Phys.\ {\bfseries #1} \ifemp{#2} #3}
\newcommand{\jp}[3]{J.\ Phys.\ {\bfseries #1} \ifemp{#2} #3}
\newcommand{\zp}[3]{Z.\ Phys.\ {\bfseries #1} \ifemp{#2} #3}
\newcommand{\ncim}[3]{Nouvo.\ Cim.\ {\bfseries #1} \ifemp{#2} #3}
\newcommand{\cqg}[3]{Class.\ Quant.\ Grav.\ {\bfseries #1} \ifemp{#2} #3}
\newcommand{\prs}[3]{Proc.\ Roy.\ Soc.\ {\bfseries #1} \ifemp{#2} #3}

\newcommand{\mybibitem}[1]{\bibitem{bib:#1} }

\end{document}